# AI-Driven Tactical Communications and Networking for Defense: A Survey and Emerging Trends

Victor Monzon Baeza, Senior Member, IEEE, Raúl Parada, Laura Concha Salor, and Carlos Monzo, Senior Member, IEEE

*Abstract*—The integration of Artificial Intelligence (AI) in military communications and networking is reshaping modern defense strategies, enhancing secure data exchange, real-time situational awareness, and autonomous decision-making. This survey explores how AI-driven technologies improve tactical communication networks, radar-based data transmission, UAV-assisted relay systems, and electronic warfare resilience. The study highlights AI applications in adaptive signal processing, multi-agent coordination for network optimization, radar-assisted target tracking, and AI-driven electronic countermeasures. Our work introduces a novel three-criteria evaluation methodology. It systematically assesses AI applications based on general system objectives, communications constraints in the military domain, and critical tactical environmental factors. We analyze key AI techniques for different types of learning applied to multi-domain network interoperability and distributed data information fusion in military operations. We also address challenges such as adversarial AI threats, the real-time adaptability of autonomous communication networks, and the limitations of current AI models under battlefield conditions. Finally, we discuss emerging trends in self-healing networks, AI-augmented decision support systems, and intelligent spectrum allocation. We provide a structured roadmap for future AI-driven defense communications and networking research.

*Index Terms*—Artificial Intelligence, Tactical Communications, Information Network, Improved Tactical Scenarios, Defense

## I. Introduction

The emergence of Artificial Intelligence (AI) into military communications and networking is transforming modern defense strategies alongside exponential data availability and computational power growth. This marks a defining moment for merging military and civil communication systems in the contemporary digital information era. AI-driven advancements enhance secure data exchange, real-time battlefield awareness, and autonomous decision-making across various tactical domains. In healthcare [1], AI enhances diagnostics and predictive analytics; autonomous vehicles and intelligent traffic systems improve safety and efficiency in transportation [2]. Similarly, AI drives advances in manufacturing through robotics, predictive maintenance, and real-time supply chain management. These achievements underscore the transformative potential of AI in civilian domains and its capacity to unlock new technological possibilities [3].

AI extends into space, enhancing satellite networks and ground communications [4], boosting resilience, autonomy, and operational superiority. Information and Communication Technology has been a fundamental pillar in pursuing safer environments in the military domain [5], [6] to achieve more innovative services. However, the same evolution has not been observed with AI. Despite the successes in civil environments, the military or tactical defense sector has been slower in adopting AI, hindered by unique challenges [7], [8]. Military operations increasingly rely on tactical communication networks, Unmanned Aerial Vehicle (UAV)-assisted relay systems, radar-based data transmission, and electronic warfare to maintain strategic superiority. AI is pivotal in optimizing these systems by improving adaptive signal processing, multi-agent coordination for network resilience, and AI-driven electronic countermeasures. However, military AI research remains fragmented despite these advances, with limited consolidation of existing methodologies and applications. Ethical concerns, such as using AI in autonomous weaponry, provoke global debates over accountability and morality. At the same time, the separation between civilian and defense communications research limits opportunities to adapt proven AI methods for tactical purposes. These issues are compounded by algorithmic biases, interpretability challenges, and regulatory hurdles, all hindering the widespread adoption of AI in tactical defense.

The hesitation to embrace AI is further reflected in the lack of comprehensive research and surveys guiding its adoption in military contexts. This fragmented landscape obscures best practices and slows progress. Unlike civilian industries, where innovation can be iteratively tested, military AI systems require rigorous validation due to the catastrophic consequences of failure, such as loss of life or escalation of conflict. To combat this issue, games or video games have served as ideal testbeds for AI research due to their characteristics that mirror real-world challenges [9]. While primarily focused on gaming environments, it offers valuable insights into evaluating and validating AI systems in controlled settings that simulate actual conditions. Games have become a key asset for advancing research in the military domain through simulators [10]; however, their application has been more prevalent in fields like autonomous driving [11], where immediate risks are critical or virtual reality [12] rather than being widely

V. Monzon Baeza, L. Concha and C. Monzo, are with Universitat Oberta de Catalunya (UOC), Barcelona, Spain. R. Parada is with Centre Tecnològic de Telecomunicacions de Catalunya (CTTC), Spain. Corresponding author: Victor Monzon Baeza (vmonzon@uoc.edu).

considered for tactical or military environments. These barriers emphasize the need for a balanced approach that fosters innovation while ensuring ethical compliance and operational reliability.

Emerging technologies like 5th Generation (5G) and Digital Twin (DT) are beginning to bridge this gap, offering new horizons for defense innovation [13]. 5G networks enable ultra-fast, low-latency communication, facilitating real-time data exchange for AI-driven systems such as autonomous vehicles and sensor networks [14], [15]. Digital twin technology, which creates virtual replicas of physical systems, enhances mission planning, predictive maintenance, and risk-free simulation capabilities [16]. DT in defense is proposed in [17]. These advancements address current challenges and create a digital ecosystem conducive to AI integration.

Leading nations already incorporate limited AI into military operations in this evolving landscape. Applications range from intelligence, surveillance, and reconnaissance Intelligence Surveillance and Reconnaissance (ISR) to logistics, command, and control systems [18] to improve national security. For instance, the United States (U.S.) Army is leveraging AI to revolutionize logistics and supply chain management, optimizing resource allocation processes [19]. Additionally, the U.S. Department of Defense's Joint All-Domain Command and Control initiative aims to connect sensors from all military branches into a unified network powered by AI, enhancing decision-making and operational efficiency [20]. These AI-enabled systems aim to augment human decision-making, manage vast amounts of data, and introduce new operational concepts like autonomous swarming to gain tactical advantages. These efforts reflect a broader trend among leading nations to incorporate AI into various military domains to enhance capabilities and maintain a strategic edge. However, challenges such as algorithmic bias and ethical risks persist, necessitating clear frameworks for responsible AI governance.

Bridging the gap between civilian and military research fosters collaboration, accelerates innovation, and leverages established expertise for defense applications. For example, concepts like swarm intelligence in autonomous systems derived from civilian robotics can provide a strategic advantage in military contexts. Additionally, advancements in AI-driven radar technology from the civilian sector can enhance tracking and detection capabilities in tactical scenarios or automatic learning incorporated into electronic warfare to improve interference detection. Therefore, conducting a comprehensive survey that consolidates the current state of AI techniques and applications already employed in the military domain is essential. This survey will serve as a foundation to unify civilian and tactical research, offering a valuable tutorial for future studies.

### A. Related works and Limitations

The application of AI in defense is not a recent development; it dates back to the 1990s in the U.S. when the first strategies for incorporating AI into this sector began to take shape [21]. However, its progress has been slow, and various studies have attempted to compile advancements in this field over the years until a resurgence of interest and breakthroughs emerged around 2021. Fig. 1 illustrates the timeline of the main contributions in the collection of works related to AI in defense.

The collection of works in Fig. 1 presents a journey through the evolving landscape of AI in military applications as a strategic, surveys, overviews, panoramic or ideas. Starting with exploring AI techniques to enhance military simulations, [21] focuses on hybrid systems and terrain analysis to include realism in training. Still, it is not until 2021 [22] that a significant leap occurs, marking a turning point where AI applications in military defense begin to demonstrate substantial and practical impact. The research progresses to broader analyses of AI's impact on military security and societal systems [22], providing an expansive view of AI's role in cybersecurity, logistics, and object detection. A shift toward specialized applications emerges with the transformative role of Deep Learning (DL) in Electronic Warfare (EW) [23], reformulating traditional problems through these advanced AI models and discussing the impact of AI on tactical autonomy, addressing the challenges of developing trustworthy, explainable AI systems for defense operations. [24] starting 2022. The focus expands to encompass modern warfare methodologies, targeting strategies, cybersecurity enhancements, and military decision-making processes [25]. In 2023, the integration of AI with robotics in defense decision-making is studied in [26]. The naval forces are the first among military branches to integrate AI into their operations, with dedicated research focusing on autonomous systems for intelligence and surveillance tasks [27]. The narrative continues in 2024 with the rise of trends in Unsupervised Learning (UL) algorithms and ML Operationss (MLOps) techniques in the defense sector, highlighting approaches to manage large, unlabeled datasets effectively in [28]. The strategic use of AI for enhancing military and economic data analysis, focusing on Large Language Models (LLMs) and their impact on defense capabilities, is analyzed in [29]. Human-AI collaboration is at the forefront of tactical mission planning and optimizing operational decision-making. This cooperation is examined within air battle management systems in [30]. A review of methods for tracking user trust and mental states during cyberattacks, focusing on AI-enabled decision-making for the Royal Canadian Navy, is scrutinized in [31]. An extensive review of emerging defense technologies, including limited AI, cyber warfare, and unmanned systems, is provided in [32]. Concluding the timeline of milestones, Neuro-Symbolic AI emerges as a transformative force, blending neural networks with symbolic reasoning to enhance military decision-making and autonomous operations [33]. This progression illustrates a dynamic evolution from foundational AI applications strategically integrated into different defense procedures.

The analysis of the existing works in this field reveals recurring gaps that highlight key challenges in the current landscape of AI applications within military contexts. Many studies, such as those focusing on military simulations [21] and EW [23], are limited to theoretical frameworks or controlled environments, lacking real-world operational insights and implementation examples. Works discussing tactical autonomy and decision-making [24], [26], [33] emphasize conceptual models but fall short of providing empirical validation through real-world case studies. Moreover, while some studies explore ethical implications and societal impacts of AI [22], [29], [34], they often lack deep technical analysis related to the practical deployment and integration of AI systems in defense operations. Research on trust and human-AI collaboration [31] focuses on theoretical models of perception without addressing the operational challenges faced in real defense environments. Several reviews [25], [28], [30] heavily rely on theoretical constructs, neglecting the development of real-time, explainable AI systems crucial for mission-critical military applications. This integration has primarily focused on decision-making, process automation, and cybersecurity. However, with its diverse missions, the tactical battlefield encompasses more than just cybersecurity. A wide range of technologies, operations, and systems have the potential to be enhanced through AI. Furthermore, this chronological timeline presents concepts in isolation, neglecting the broader military perspectives that span all branches of the armed forces (navy, land, air, and space). Moreover, studies have focused on specific techniques without offering a comprehensive overview of the entire range of AI capabilities. Current overviews take a more strategic approach, addressing specific challenges of AI in defense and focusing on particular case-based needs. These gaps underscore the need for future survey research to incorporate more empirical studies, operational data, and comprehensive analyses that bridge the divide between conceptual models and real-world military implementations. As we will explore in this survey, these aspects have historically been considered separately. Therefore, it is vital to integrate all of these elements into a single foundational work.

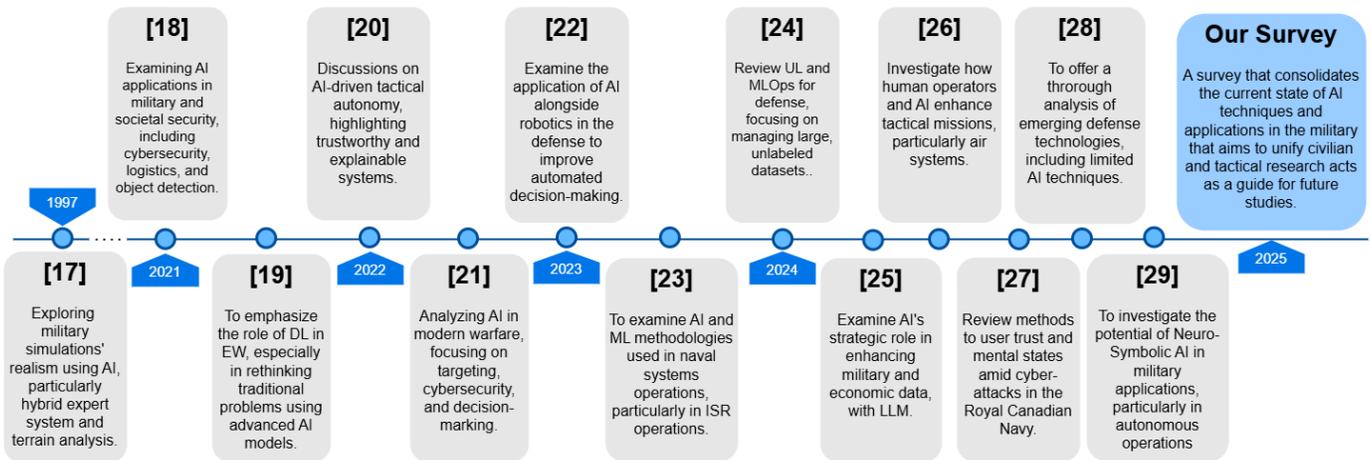

Fig. 1: Chronological timeline of existing research leading up to our survey.

Table I compares key aspects extracted from the existing literature, as identified through the chronological analysis, and contrasts them with the broader and more integrative approach adopted in this survey. The comparison between our survey and the existing papers reveals certain limitations for the state-of-art versus key advantages in our work, highlighting the necessity for a more current, comprehensive, and integrative study. While the chronological papers focus on specific AI applications or isolated military sectors, our paper offers an extensive review covering seven key military systems, providing a broader, more unified scope. Unlike previous studies, which often maintain a separation between civilian and defense research, our survey effectively bridges the gap, showcasing how civilian AI technologies can be adapted for military applications.

One of the most significant differentiators is the presence of empirical data. Whereas the reviewed papers frequently lack real-world validation, relying heavily on theoretical models, our survey incorporates analysis of real-world projects, enhancing the practical relevance of our findings. Additionally, while ethical and legal discussions are only briefly touched upon in other papers, we provide an in-depth analysis of these issues, including critical standardization challenges in military contexts.

From an operational perspective, our paper moves beyond theoretical concepts to present a detailed evaluation of military system performance backed by statistical insights. Unlike many previous works, which often provide limited operational data, our survey identifies technological gaps and offers actionable solutions rather than merely recognizing them. Furthermore, our emphasis on tactical environments is more comprehensive, addressing critical operational needs often neglected in other studies. By integrating statistical insights into AI research trends, we enhance the value of our work, providing quantifiable evidence to support our analyses.

Lastly, another strength is the future-oriented perspec-tive. While some chronological papers briefly mention future work that is now outdated, our paper provides a comprehensive outlook on emerging AI trends. This is complemented by an accessible framework designed to engage diverse audiences, including military professionals, researchers, and non-specialists—a level of inclusivity rarely found in prior surveys.

TABLE I: Comparison of our work to relevant existing works in the field.

| Aspect | Our survey | Related works: [21]–[33] |
|---|---|---|
| Comprehensive Scope | Extensive, covering 6 key military technologies and systems | Focused on specific AI applications or sectors |
| Civilian-Military Integration | Bridges civilian AI with military applications | Limited integration between civilian and military contexts |
| Empirical Data | Includes analysis of real-world projects | Lacks extensive empirical and real data |
| Ethical and Legal Analysis | In-depth ethical, legal, and standardization discussions | Surface-level ethical discussions. Lack of legal and standardization perspectives |
| Operational Insights | Detailed evaluation of military system | Theoretical insights |
| Technological Gap Identification | Identifies gaps and proposes solutions | Limited identification of technological gaps |
| Accessibility for Diverse Audiences | Designed for both military and civilian researchers | Primarily academic or defense-specific audiences |
| Focus on Tactical Environments | A strong focus on tactical applications | Focused on constrained tactical environments. |
| Statistical Insights | Provides statistical insights into AI research trends | Rarely includes statistical analysis |
| Future Perspectives | Forward-looking analysis on AI evolution | It does not include current AI versions such as Gen-AI or LLMs. |

In conclusion, regarding related works, no existing survey has provided a comprehensive assessment encompassing all the key elements necessary to evaluate military communications and networks within a tactical environment. Previous works have primarily focused on isolated technological aspects, lacking a holistic approach that integrates technical and application-oriented perspectives. This survey aims to fill that gap by analyzing diverse AI techniques and considering critical military factors such as operational requirements, strategic deployment, and multi-domain integration (land, sea, air, and space). By connecting these domains, this work seeks to expand the scope of the investigation, making it accessible to a wider range of researchers and encouraging cross-sector innovation in AI for defense applications.

### B. Scope and Contributions

This paper represents the first comprehensive survey of tactical communications and networking aided by AI, analyzing the state of research on AI applied to military communications and networking technologies. Despite the rapid advancements in AI, little effort has been made to evaluate its application within the military domain systematically. Our review seeks to bridge this gap by providing a detailed overview of current developments, challenges, and opportunities in this field of tactical communications and networks.

We emphasize the potential of leveraging advanced civilian AI techniques that could be adapted for military communications and networking use but have yet to see widespread implementation. This gap is often due to limited accessibility or the specialized nature of military communications. By synthesizing civilian and military research insights, we aim to facilitate a deeper understanding of these technologies and their potential cross-domain applications.

The scope of our work is an approach designed to be accessible to a diverse audience, including experts from both civilian and military sectors and non-specialists interested in the intersection of AI and defense. We provide clear and structured descriptions of the main systems, focusing on making complex technologies understandable and relatable. To achieve this, we classify and analyze existing research, technologies, and applications, offering a comprehensive perspective that includes assessing significant projects in the defense industry. Our analysis goes beyond technical considerations to address critical issues, such as the ethical and legal implications of deploying AI in defense systems. Finally, we present forward-looking perspectives on how AI could shape the future of military and tactical operations. This includes identifying unresolved challenges and exploring the potential of emerging technologies to transform defense strategies. By doing so, we aim to establish a foundation for further academic and industrial exploration of AI in military contexts.

We tackle this by making the following contributions.

1) First Comprehensive Survey in AI-driven military or tactical communications and networks.
2) Bridging Civilian and Military Communications and Networks: The review highlights advanced civilian AI technologies with untapped military potential.
3) Accessible Framework for Diverse Audiences: We offer clear descriptions of military communication and networking technologies and systems, making our analysis relevant for non-specialist audiences.
4) State of Research in AI for tactical networks: It establishes the current status of AI research in military contexts, presenting a consolidated resource for the academic and defense communities.
5) Overview of Relevant Industry Projects: The review synthesizes key defense industry projects, highlighting real-world AI applications in military systems.
6) Exploration of Ethical and Legal Challenges: The analysis delves into critical issues such as the ethical, legal, and societal challenges posed by adopting AI

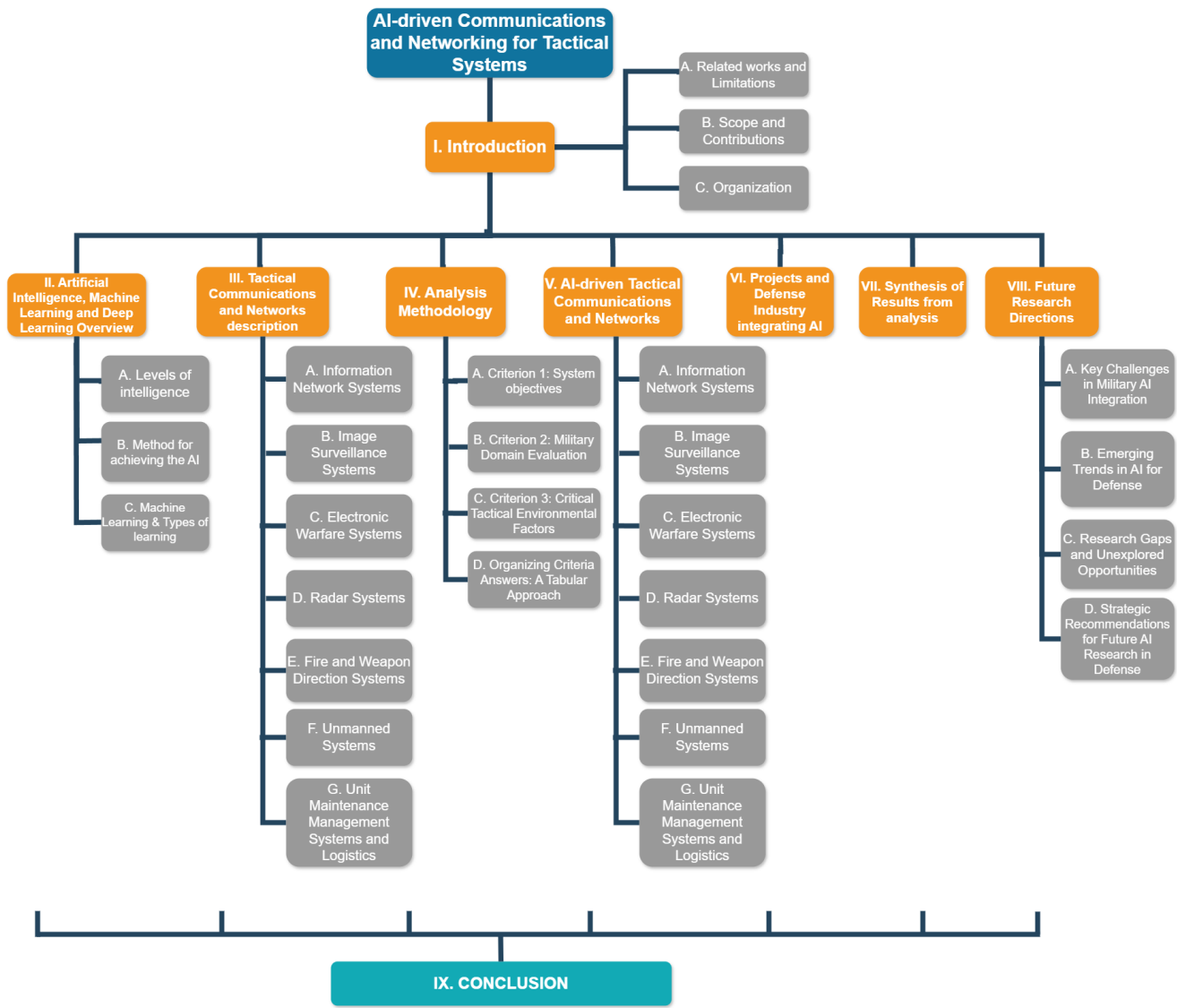

Fig. 2: Structure of this survey.

in defense technologies.

7) Future Perspectives in Tactical AI: It highlights unresolved issues and examines technologies that could transform military operations through AI innovation.

## C. Organization

An overview of the organization and structure of this paper is illustrated in Fig. 2. After an introduction to understanding the necessity and motivation of this study, we outline the following sections. Section II provides an overview of AI concepts, covering levels of intelligence, methods to achieve AI, and Machine Learning (ML) paradigms relevant to military applications. Section III introduces the seven core military systems that are the foundation for AI integration, including tactical information systems, surveillance, electronic warfare, radar, fire control, unmanned systems, and logistics. Section IV defines the three evaluation criteria for assessing AI's impact on these technologies. Section V revisits the seven military systems, analyzing how AI enhances decision-making, automation, and operational effectiveness in each domain. Section VI explores real-world defense initiatives and industrial projects actively deploying AI in military operations. Section VII consolidates insights from the previous sections, summarizing trends, key findings, and cross-domain implications of AI applications in defense. Section VIII highlights key challenges, emerging trends, research gaps, and strategic recommendations necessary to advance AI in military contexts. Finally, Section IX concludes this survey, emphasizing the importance of AI in defense, its potential impact, and the need for responsible, secure, and interoperable AI deployment.

## II. Overview of Learning Types

AI is a technology that offers new projects in R+D+i and has brought a completely different perspective to wars and military strategies. It provides a new tool in knowledge engineering, Big Data learning engineering, and the simulation of tactical strategies and environments. It could be said that the impact it has had on military theories and the arts of war is so relevant that it has led to new ways and means of conceiving it [35].

This technology tests aspects such as the science of thought, cognitive science, and information science, among others. The exponential growth in the number of projects carried out with AI and the results clearly show that this resource could change the future as we know it today.

New research, visions, and potentialities emerge daily in the military, confirming that the arms race will fundamentally be based on AI. It is envisaged that the machine will observe, orient itself, make decisions, and act accordingly [35]. An army with multiple or even infinite intelligent brains acting autonomously on the battlefield or assisting the commander in decision-making will make all the difference in confrontations.

### A. Levels of intelligence

Knowing that there are three levels of intelligence helps us understand AI and ML. Fig. 3 shows below where these levels are represented and then proceeds to their explanation.

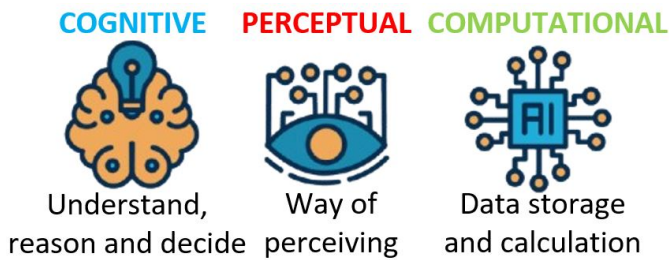

Fig. 3: Intelligence levels

In the first place, there would be cognitive intelligence (blue), whose purpose is to understand, reason, and decide. It is based on neuro-linguistic programming, and in some scenarios, it has managed to exceed the capabilities of human beings. For example, the AlphaStar computer program eliminated professional players from the strategy game StarCraft II in real time [36].

On the other hand, there would be perceptual intelligence (red), which is precisely related to the way of perceiving, a perception based on the senses with a subjective connotation. Except for taste, which, to the best of the author's knowledge, no publications have been found on sight, touch, hearing, and even, recently, smell are immersed in this level. Deep Neural Networks (DNN) and Big Data have evolved enormously in this concept, surpassing even human perception. Examples of these developments are image recognition programs such as Google Lens [37] or facial recognition programs with emotion detection such as Google Vision API [37], speech recognition systems such as Windows Speech Recognition or Dragon Naturally Speaking [38] or language translation systems with error rates around 5% [39].

Finally, there would be computational intelligence (green) responsible for data storage and calculation. Logically, the greater the amount of data that can be stored and the greater the speed with which it can be calculated or worked, the better. Initially, this type of intelligence was mainly based on arithmetic and logical tasks. Still, today, it includes learning, adaptation, and fuzzy logic, which allows, in a certain way, to conceive of this intelligence. Currently, computers have far surpassed humans in this aspect. For example, one could mention the development of the MareNostrum 5 computer located in Barcelona, which will reach a power of 314 petaflops [40]. Other examples of powerful supercomputers are IBM's Summit with 200 petaflops [41] or China's Tianhe-3 with a peak of 1.3 exaflops [42].

### B. Method for achieving the AI

The method that AI uses to achieve this intelligence is based on the search for the solution, the inference of knowledge, and ML as developed in the reference [35]. Fig. 4 illustrates this model, which will be explained later. When we talk about "looking for the solution," we mean the search process composed of algorithms, rules, and methods used whose purpose is to encompass the resolution of the problem from the level of representation. At this point in the search for a solution, the decision factor is kept out of the equation since it studies the different states of the problem and the transitions between them to find the best way from the initial state to the final state. In other words, in this section, the purpose is to find feasible solutions or paths where all possible routes have previously had to be studied. It will be decided at another point in the AI which path is the best since there are many factors to consider, such as energy consumption, duration, danger, distance, and success rate.

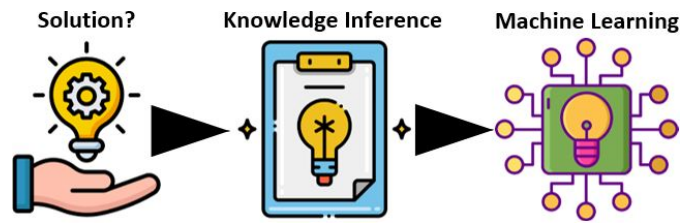

Fig. 4: Method to use AI

On the other hand, there would be the inference of knowledge whose main objective, broadly speaking, would be to represent formal knowledge in such a way as to facilitate reasoning. That is, to draw valid premises and conclusions that represent formal knowledge and can be used to think and solve problems in computers or intelligent systems by simulating modes of reasoning and control strategies. In this abstraction of conclusions, a significant problem arises: the uncertainty that occurs when the logical relationship between the premises and the conclusions is not found or does not exist at all. To solve these questions, the system relies on an inference engine whose function is to try to reason, search for the information in the available knowledge base, associate it with the elements of a database, and create or expose new ideas guided by some control strategy. A reasonable inference engine should stand out for its efficient search and matching mechanism, controllability, observable representation of cause and effect, and heuristic quality.

The inference of knowledge from AI systems creates two relevant problems. The inference method involves analyzing logical relationships and evaluating the reliability of these connections. Conversely, the control strategy minimizes and streamlines the search efforts required for timely answers. To implement this control strategy, one can use rule learning—an intelligent program with significant knowledge and expertise—or fuzzy logic, mainly when the descriptive model is uncertain or shows limited linearity.

Finally, the third technology AI uses is ML, which is the science that studies computer learning from available data. It mainly consists of identifying and understanding patterns and patterns from a large volume of samples to predict or solve practical problems.

### C. Machine Learning & Types of learning

There is an extensive bibliography on ML and the types of learning used. According to the information available in references [43] and [44], Fig. 5 presents a summary of the types of learning along with the algorithms most commonly used in each of them and then explains them in a very summarized way.

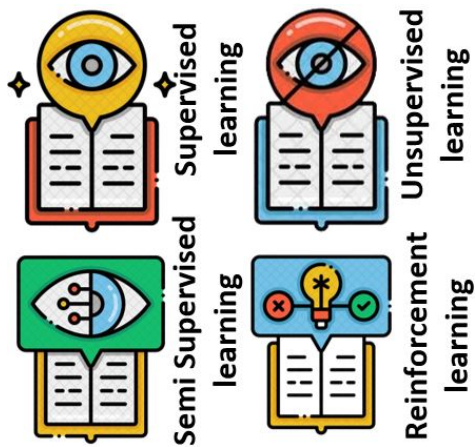

Fig. 5: Type of ML

- Supervised Learning (SL): The training data supplied to the algorithm includes the desired solutions; it is labeled. A set of samples from known categories is used. An example would be the classification of the email spam filter or the prediction of a target numerical value given a set of characteristics, such as vehicle valuation. It can also predict or analyze data from the Internet of Things (IoT), social networks, or facial recognition techniques.

- Unsupervised Learning (UL): Training data is not labeled, and the system attempts to learn without a teacher through groupings, visualizations, dimensionality reductions, or association rules. It's used a lot in data mining to figure something out. For example, in the search for a person in a plot of different images, in a visualization algorithm for graphical representation of unlabeled data that facilitates traceability when grouping them, in a hierarchical grouping or clusters of visitors to a blog for detecting similar groups of users, simplifying data by merging several correlated features into one, detecting anomalies such as defects in a production chain, and learning data association rules to discover the relationship between them.

- Semi-supervised Learning (SSL): Used when much of the data being handled is unlabeled, some partially labeled, and a few labeled. For example, we have many photos, and the system identifies that person 1 appears in photos A, C, and D (unsupervised part) and then tags them with their name in one of them. The system automatically tags them in all of them, thus facilitating their search.

- Reinforcement Learning (RL): This type of technology is based on observation, choosing a policy of action, and maximizing rewards or penalties. The vast majority of AI systems specialized in gaming use this approach, such as AlphaGo.

On the other hand, algorithms can also be classified according to the availability of the data, the generalization of the data, or the number of layers needed to learn.

- Batch learning: It is a system that learns using all available data, but always from the beginning. If a system is trained with 2,000 samples, an algorithm is generated. If you wanted to include the information from 500 more samples, you would have to start from the beginning by retraining the system with 2,500 samples. It usually needs a lot of computing resources and time and is done offline.

- Online or incremental learning: New data is injected into the system to add the learning to the previous one so that new learning occurs immediately. The main advantages are memory saving, since once the system has been trained with data, it would no longer need to be stored, and the speed of adaptation to changing data. On the contrary, it has the disadvantage that if the new data provided to the training is erroneous (deteriorated sensor), the system's performance will worsen, and it is essential to revert to a previous state of learning. It is used in the stock market.

- Instance-based learning: The system learns from memorized examples and then discerns new incoming cases based on a measure of similarity. It is often used in the anti-spam filters of emails whose similarity measure could be the number of words in 2 emails.

- Model-based learning: The steps typically are: study the data, select a model from a set of examples, train the system until the model's parameters minimize a cost function, and finally apply the algorithm to make predictions about new events.

- Deep learning (DL): It is characterized by the fact that in its architecture, there is more than one level or layer where the parameters learn from the results of the preceding layers. They never learn directly from the characteristics of the data samples, which are located in the first layer. In addition, each layer can use its type of ML, although they are usually based on Multi-Layer Neural Networks (MNN). This type of learning attempts to mimic the human brain's mechanism for interpreting data such as images, sounds, and texts.

## III. Tactical Communications and Networks description

Before analyzing AI integration, it is essential to establish a foundation by examining the core military or tactical communications and networking technologies that serve as the backbone of modern defense operations. This section outlines seven key technologies related to tactical communication systems that have shaped military capabilities. It provides the context for understanding their potential enhancements through AI in later sections. According to Fig. 6, these technological systems are Tactical Information Networks Systems, Image Surveillance Systems, Electronic Warfare Systems, Radar, Fire and Weapon Direction Systems, Unmanned Systems, Unit Maintenance Management Systems, and Logistics.

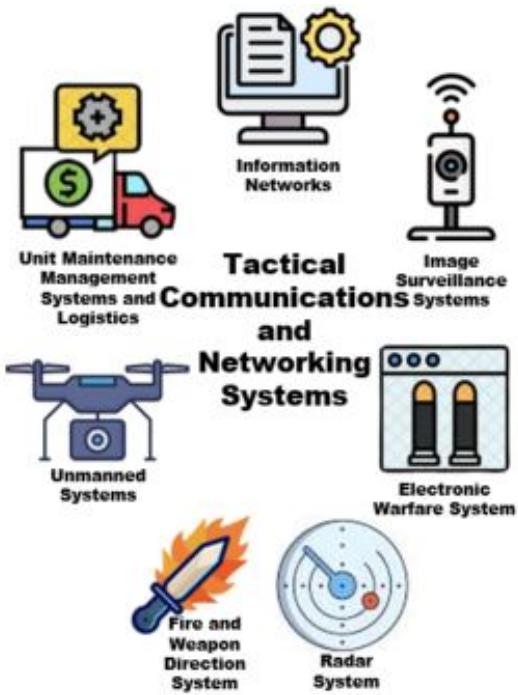

Fig. 6: Core Military Technological Systems.

### A. Information Network Systems

Information systems are tools designed to collect, process, analyze, and distribute relevant information in real-time or near real-time within a military or high-security context. Their primary goal is to provide commanders and operational units with accurate data to make decisions during tactical operations in scenarios where speed and precision are critical. Information is power [45], and having access to practical, accurate, and truthful information is the most critical factor when making any decision, whether it involves establishing a business plan or defining a military strategy. From the beginning, there have been acoustic or visual signals providing commanders with battlefield updates, progressing through horseback messengers with sealed letters, the invention of terrestrial radio communications, satellite links, and many other innovations up to the present situation.

The data collection process integrates multiple sources of information, such as sensors, drones, satellites, ground communications, and portable devices. The information processing component analyzes large volumes of data in real-time to identify patterns, threats, or environmental changes. Information is distributed through secure and reliable communication networks and protocols. For instance, in the Navy, the LINPRO tactical network processor [46] manages real-time information exchange between networks connected via protocols such as Link 11, Link 16, Link 22, Variable Message Format (VMF), or Joint Range Extension Application Protocol (JREAP). These capabilities allow for receiving and transmitting tactical information over long distances, often through satellite networks, while adhering to North Atlantic Treaty Organization (NATO) standards like STANAG. A robust tactical information network system must include features of scalability and adaptability to diverse operational scenarios, ranging from ground operations to aerial or naval combat.

### B. Image Surveillance Systems

An image surveillance system in a military context is a network technology designed to monitor, capture, and analyze visual data from strategic areas to enhance situational awareness and decision-making. These systems use advanced cameras, sensors, drones, and satellite imaging to provide real-time or near-real-time visual intelligence, identify threats, track movements, and ensure perimeter security.

It was around 1969 when the first domestic closed-circuit television (CCTV) system was recorded [47], although similar systems had already been used in military projects years earlier. Technological advancements since then have been remarkable, improving image quality, incorporating infrared vision, enabling thermal data options, operating with servers and digital recordings, implementing facial recognition, motion-triggered pixel activation, vehicle license plate recognition, video tracking, and much more, as noted in [48].

All this technology has brought significant changes to surveillance systems, as the types of sensors used for data capture are highly diverse [49], and data recognition and processing software offer extensive capabilities.

## C. Electronic Warfare Systems

EW involves technological and electronic activities aimed at detecting, exploiting, disrupting, or denying the adversary's hostile use of all energy spectrums—such as the electromagnetic spectrum—while ensuring its continued use for its own benefit [50]. Notably, this type of warfare impacts radio communication transmissions, surveillance radar systems, and electronic fire control systems.

Due to the diversity of scenarios and systems, EW can be broadly divided into three main types of measures or fundamental components:

- Electronic Support Measures (ESM):These involve actions taken to search for, intercept, identify, or locate sources of emitted electromagnetic energy to gain immediate recognition of potential threats.

- Electronic Countermeasures (ECM): These consist of actions aimed at denying or reducing the enemy's use of the electromagnetic spectrum. This includes jamming, deception, and various decoys used for missile defense.

- Electronic Protective Measures (EPM): These involve measures taken to ensure the reliable use of the electromagnetic spectrum for friendly forces-for instance, fire-control radars are equipped with frequency-hopping agility.

The origins of EW systems can be traced back to the WLR-1, developed by the U.S. in the 1950s. By the 1970s, Italy introduced the Beta Mk1000 system. Spain began developing its ESM system with the DENEB program in the 1980s. Over time, the Spanish Navy has equipped its vessels with systems such as ALDEBARAN and REGULUS on F100-class frigates and the RIGEL system on the LHD Juan Carlos I and BAM ships. It is advancing with Indra's development of the RIGEL i110 and REGULUS i110 for the next-generation F110-class frigates. Additional insights into the evolution of these systems within the Spanish Navy can be found in references [51] and [52].

Most ESM systems have similar architectures. They typically consist of an operator console equipped with a loaded signal library, a signal processing rack to analyze intercepted signals and measure delays, and a module installed on the superstructure that performs initial signal filtering to determine direction. Additionally, the system includes three antennas mounted on the superstructure: one omnidirectional antenna and two directional antennas positioned on either side of the platform to capture energy from those sectors.

## D. Radar Systems

A radar in military systems is a critical electronic device that detects, tracks and identifies objects at a distance by transmitting radio waves and analyzing the reflected signals. It plays a vital role in various military operations, including surveillance, target acquisition, missile defense, and navigation. Radars can be classified based on their function, such as search, tracking, or fire control, and are essential for providing situational awareness in complex and dynamic environments. These systems help military forces detect threats, monitor movements, and enhance the effectiveness of strategic operations, often operating in challenging conditions like adverse weather or low visibility.

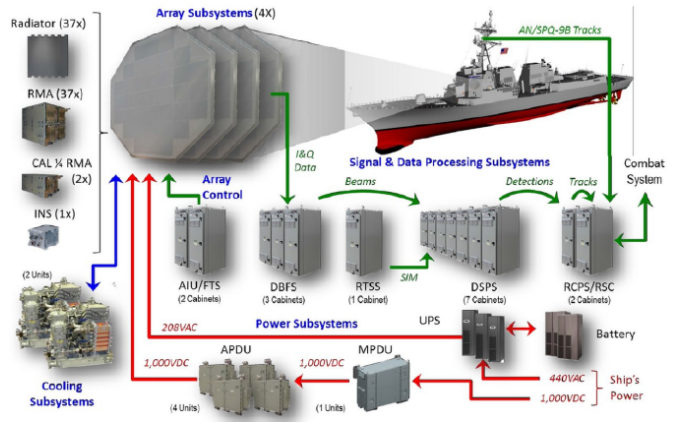

Fig. 7: General description of AN/SPY-6 system [56].

The British developed the first military radar in 1934, and significant advancements have been made in radar technology from this date, as outlined in [53]. The basic functional design remains consistent, but considerable progress has been achieved in areas such as antenna types, signal processors, transmitted radio frequency power levels, oscillators, and classifications by technology (e.g., primary, secondary, pulsed, continuous wave) or application (e.g., air traffic control, meteorology, navigation, tracking), as summarized in [54].

One of the most advanced and versatile radars in the military domain is the SPY-7 developed by the American company Lockheed Martin, as presented in [55]. This S-band digital radar, built using gallium nitride solid-state technology, features modular and scalable software-defined architecture. It can detect, track, and engage sophisticated ballistic missile threats, even simultaneously managing multiple targets. Moreover, it is interoperable with most existing defense radars and platforms. Since it is still under development, limited literature on the SPY-7 exists. However, some insights can be drawn from its predecessor, the SPY-6, whose modular composition is illustrated in Fig. 7.

The SPY-6 [56] is an Active Electronically Scanned Array (AESA) radar consisting of three main components: an S-band Air and Missile Defense Radar (AMDR) that provides volume search, missile tracking, and discrimination; an X-band AMDR, which provides horizon and surface search, precision tracking, and terminal illumination; and an AMDR Radar Suite Controller that coordinates and integrates both radars. This AMDR is the first radar built using 2' x 2' x 2' Radar Modular Assemble (RMA) building blocks, allowing for scalability and utilizing Gallium Nitride (GaN) in its construction to require less power and enhance cooling efficiency.

This radar also supports digital beamforming, which enables more accurate tracking, greater range, and 30 times the sensitivity of other radars. Additionally, the SPY-6 has offensive capabilities, including conducting electronic attacks using its AESA antenna. It can target airborne and surface targets using tightly focused, high-power radio wave beams that could potentially blind adversary assets.

Situational awareness, or understanding the environment in a military context, is crucial for decision-making in tactical operations. The radar system plays a key role in providing this information. In the recent case of [57], an over-the-horizon radar system for surveillance and knowledge networks is presented, which maximizes detection accuracy and the characteristics of the data. However, the optimization and detection process excludes AI methods.

### E. Fire and Weapon Direction Systems

A Fire and Weapon Direction System (FWDS) is a critical component in military operations, designed to provide accurate targeting, control, and guidance for weapon systems during combat. It integrates data from various sensors, including radars, electro-optical devices, and targeting systems, to calculate the optimal firing solution. The FWDS helps direct weaponry, such as guns, missiles, or other armaments, ensuring precise and effective engagement of targets. By processing real-time information, the system enables commanders to make quick, informed decisions and optimize available firepower, enhancing operational efficiency and combat effectiveness in dynamic environments.

With the advent of heavy artillery, documents were created that correlated the amount of gunpowder, the weight of the projectile, the angle of the cannon, and the range achieved by the projectile based on these factors. These documents were highly valuable, and when combined with the skill of artillery personnel, they significantly increased the likelihood of successfully hitting the target, as described in [58].

Today, FWDS have advanced significantly, and any system associated with a weapon of considerable caliber will typically include, at a minimum, an infrared optical sensor, a daytime optical sensor, and a laser rangefinder. For example, according to the ATLAS project, which will be discussed in the next section, tanks will be equipped with image sensors across various wavelengths, including visible, Near Infrared (NIR), Short-Wave Infrared (SWIR), Mid-Wave Infrared (MWIR), and Long-Wave Infrared (LWIR), along with a laser rangefinder, all of which will feature continuous 360° rotation for target acquisition and identification, as well as for use in fire control.

In other units, such as warships equipped with 76 mm caliber guns or similar, capable of engaging targets at distances greater than 15 km, the fire control system includes a continuous-wave radar for target tracking to ensure accurate targeting. At such long ranges, factors such as rain, cloud cover, haze, or any intermediate obstacles between the ship and the target may hinder image-based tracking but would not affect radar tracking capabilities.

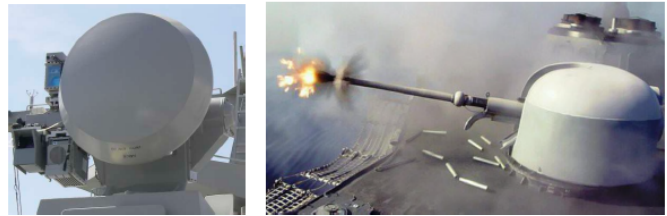

Fig. 8: Pedestal radar with DORNA-2 fire control sensors [59] and Oto-Melara 76mm cannon [60].

Fig. 8 illustrates the pedestal of the DORNA-2 fire control system installed on Spanish warships [59], which includes radar, laser rangefinder, Charge-Coupled Device (CCD) camera, and Infra Red (IR) camera, alongside the Oto-Melara 76mm gun in the process of firing [60].

Fire control and weapon systems primarily differentiate military units from civilian ones. A warship and a fishing boat, or a military aircraft and a passenger plane, are almost the same systems. Still, the installation of fire control, weapons, and electronic warfare systems, broadly speaking, have a decisive influence on the conceptual distinction.

### F. Unmanned Systems

Unmanned Systems (US) refer to vehicles, aircraft, or vessels that operate autonomously or remotely without a human onboard, both in military and civil contexts. Unmanned vehicles can be aerial (UAV, although if the control and communication system are included, it would be a Unmanned Aerial System (UAS), commonly called drone), maritime (Unmanned Ship Vehicle (USV)), or ground-based (Unmanned Ground Vehicle (UGV)). Within this classification, a second subdivision can be made according to the type of mission assigned (surveillance, attack, suicide, etc.) or the type of technology applied (remote control, autonomous, fixed-wing, rotary-wing, etc.). Given that UASs have been making headlines in recent years due to technological advancements and their military use, it seems appropriate to focus this study on them. A diagram of drone terminologies is presented to aid in understanding in Fig. 9.

Unmanned aviation is as old as manned aviation. Still, in recent years, it has undergone the most significant evolution, mainly driven by new technologies, new materials, new energy storage systems, and the new roles these devices have played in the military world, according to reference [61]. Fig. 10 shows some examples of UASs, some as small as the palm of a hand and others with dimensions similar to those of a fighter jet. The most notable projects are the Black Hornet 3 mini-drone from FLIR Systems, for its surveillance capabilities, lightweight, and high-resolution camera, as described in [62]; the Turkish armed drone SONGAR, for its integrated 5.56 mm caliber weapon and its infantry support; and the American Predator C Avenger drone, for its reliability and its payload capacity for weaponry.

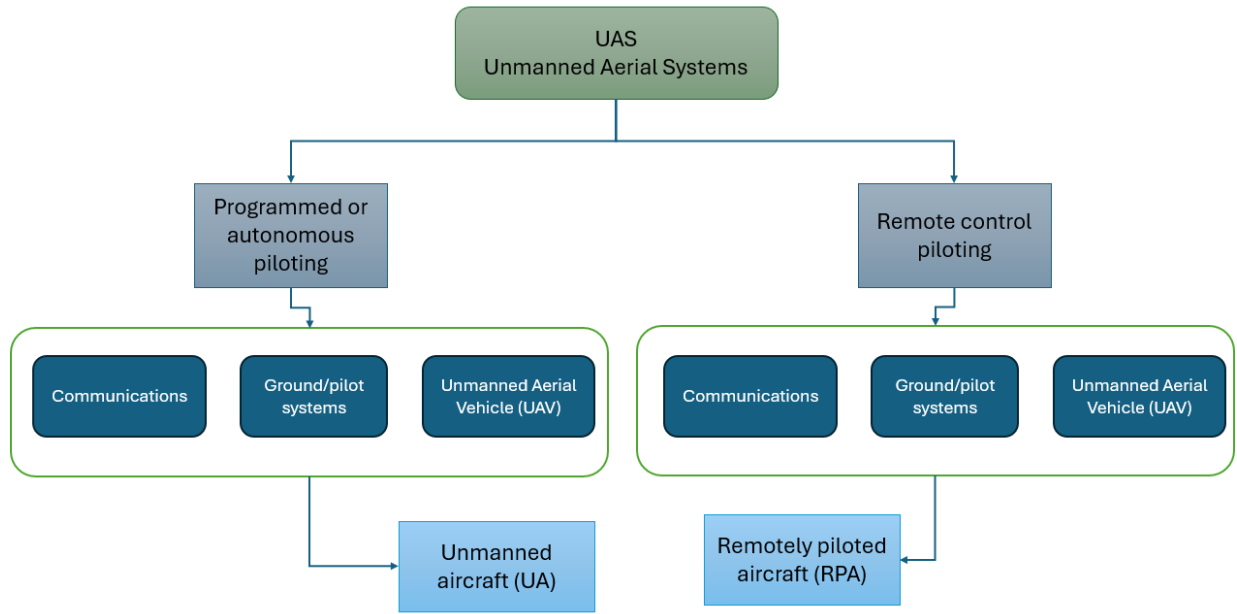

Fig. 9: US Terminology

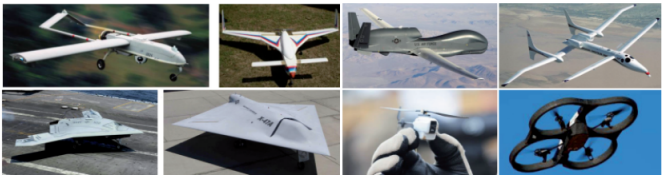

Fig. 10: Examples of UAVs

## IV. Analysis Methodology

The methodology applied in this survey to conduct the investigation is based on an analysis composed of three criteria:

- Criterion 1: Covers general data from the selected references.
- Criterion 2: Evaluate the notable and decisive factors relevant to the application in a military domain.
- Criterion 3: Assesses the critical tactical environmental factors affected by the application.

Each of the five technological systems presented in the previous section will be analyzed using this three-criteria approach through a series of tables. Consequently, each system's analysis consists of three dedicated tables, one for each criterion. Combining these three branches creates a comprehensive evaluation framework.

### A. Criterion 1: System objectives

General and foundational information is provided under this criterion, which is essential for any study. Fig. 11 visually represents the concept map around "Criterion 1" that will later be used to define this criterion in detail.

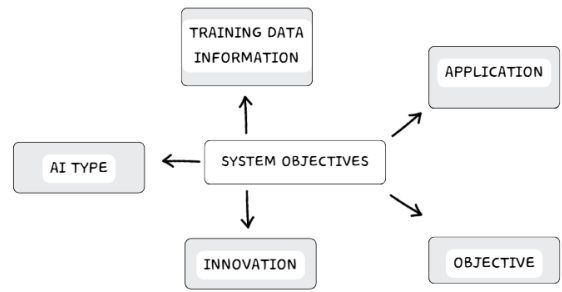

Fig. 11: System objectives of the references.

These concepts classify the references under study as follows:

- Application: It aligns with the reference title and relates to this publication's purpose, answering what is intended to be achieved.
- Objective: implemented advantages, improvements, and procedures, detailing the steps to enable their application in tactical communication and networks.
- Innovation: It is helpful to determine whether the reference introduces any concept that hasn't been seen before or is seldom used, which might be worth noting.
- AI Type: Identify the type of learning model and the algorithm utilized. Adapting a commercial design for a defense system differs significantly when the developed algorithm operates offline and is non-incremental.

- Training Data Information: Understanding the training data is crucial, as it can be inadequate, gathered under controlled conditions, or specific to a scenario. This is vital for supervised algorithms since the training data's characteristics can limit the system's scope and applicability.

## B. Criterion 2: Military Domain Evaluation

The decisive factors, highlighted as "Criterion 2", require special attention when implementing a new technique in a system, as they help identify the algorithms' potential or limitations. Based on the literature review and assimilation of the data presented in the state of the art, the necessary background has been acquired to propose the observation of factors illustrated in Fig. 12. These decisive factors must be considered when applying AI technology in military tactical environments:

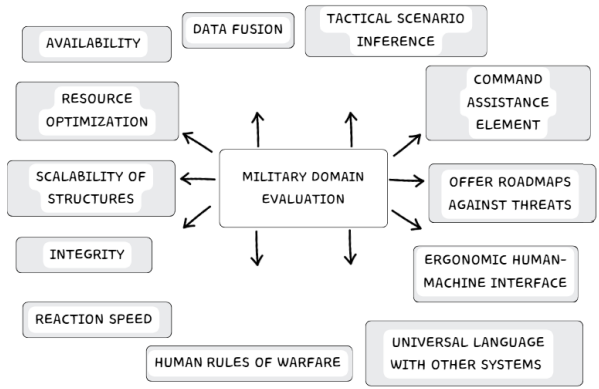

Fig. 12: Key factors in applying AI in Military Domain Evaluation.

- Data Fusion: These systems merge information and data from different sources and can process them comprehensively to obtain an accurate and reliable description of the environment. They can describe aspects of a target (speed, heading, size, armor...) or an event (involved personnel, security perimeter, topography, hostile areas...).
- Tactical Scenario Inference: The system must perceive and understand scenario elements, their spatial-temporal placement, and environmental intent. Modern warfare relies on integrated joint combat, coordinating soldiers, drones, tanks, aircraft, ships, and satellites. Combat is multidimensional, requiring commanders to access real-time battlefield data. Effective information processing and distribution across systems is crucial for operational success.
- Command Assistance Element: The decision support system consists of structured and unstructured components. The structured part involves human-machine interaction and data processing, while the unstructured part deals with uncertain, complex scenarios where traditional models cannot represent knowledge. Intelligent systems assist by analyzing warfare models qualitatively. As decision complexity grows, commanders will face an increasing gap between available information and their choices.

- Offer Roadmaps against Threats: AI assists in threat and obstacle avoidance, ensuring fast, efficient path selection. Current methods, such as genetic algorithms and dynamic planning, face challenges when extended to 3D scenarios, as large datasets slow convergence.
- Ergonomic Human-Machine Interface: Command interfaces should deliver timely, intuitive visual information over complex data tables, enhancing operator comprehension and decision-making.
- Universal Language with Other Systems: Unlike human languages in NATO operations, AI-driven machine language ensures seamless interoperability, reducing misinterpretations in multi-system environments.
- Human Rules of Warfare: While AI can enhance autonomy in tactical systems, critical war-related decisions must remain human-controlled, preventing reliance on purely rule-based learning.
- Availability: AI must ensure continuous system uptime and real-time data access; otherwise, its integration adds no operational advantage.
- Resource Optimization: Multi-dimensional combat generates vast data inputs, so AI must balance detail and efficiency to avoid unnecessary computational overload. For example, it would not make sense to go into the maximum detail of a war scenario, training the system with millions of variables unless necessary, as this would slow down the system and increase resource consumption.
- Scalability of Structures: As warfare evolves, AI systems must adapt, integrating new actors, data, and strategies without performance degradation.
- Integrity: AI must detect data manipulation attempts, ensuring consistency, validity, and security during training and operation.
- Reaction speed: AI's effectiveness in military operations depends on real-time decision-making; delayed responses negate its tactical advantage. The application of AI in military environments is justified if it improves the reaction speed of the command in response to a threat.

## C. Criterion 3: Critical Tactical Environmental Factors

When applying AI to military systems, assessing its impact on defense, particularly communications and networks, is crucial. Additionally, exploring new applications can reveal enhancements beyond a reference's initial focus.

Fig. 13 illustrates military areas impacted by AI, including potential improvements identified through synergy and direct consequences for communication systems and networks.

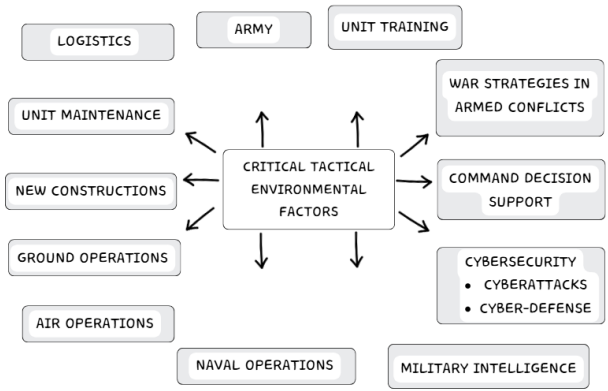

Fig. 13: Key elements to evaluate Critical Tactical Environmental Factors.

- Army: The General Directorate of Armament and Material, under the Secretary of State for Defense, oversees defense projects, aligning military needs with broader defense policies. The army defines technical and operational standards for new systems across air, naval, and ground domains.
- War Strategies in Armed Conflicts: War strategy involves planning military campaigns and troop movements to defeat adversaries. According to International Humanitarian Law (IHL), armed conflict refers to large-scale confrontations causing destruction. This study examines how ML enhances military strategies when integrated into tactical systems.
- Command Decision Support: In potential conflicts, decision-making relies on a secure, collaborative power structure for effective horizontal and vertical communication. AI-driven systems enhance real-time data processing, ensuring timely and accurate decisions.
- Cybersecurity- cyberattacks and cyber- defense: Recognized as the fifth military domain at the 2016 NATO summit, cyberspace spans battlefield sensors to Command and Control (C2) networks [63], [64]. Traditional security measures (e.g., antivirus, firewalls) now integrate ML for intrusion detection, network access control, and data protection [65]. However, ML is dual-use, serving both cyber defense and offensive cyberattacks, while also being vulnerable to adversarial manipulation.
- Military Intelligence: AI processes vast military datasets (structured, semi-structured, and unstructured) to extract actionable insights. ML clusters intelligence from messages, identity records, and communications to detect military patterns. In aerial and space surveillance, computer vision enables target identification and tracking from satellite and UAV imagery.
- New Constructions: AI applications in warfare predictions may drive adversaries to develop unforeseen strategies and structures, exploiting weaknesses in trained models. AI also enhances manufacturing processes, such as DT simulations for improving production efficiency.
- Air Operations: Conducted in the aerial domain, these operations involve highly mobile and flexible units for threat deterrence, rapid deployments, and strategic positioning.
- Ground Operations: The primary domain of human activity, hosting key political, economic, and strategic centers [66]. AI-driven analysis supports combat, defense, and stabilization efforts, securing military advantages while fostering conflict resolution strategies.
- Naval Operations: Naval forces provide mobility, availability, and interoperability [67]. Their planning and execution emphasize flexibility, goal alignment, security, and efficiency across various operational levels.
- Logistics: Military logistics involve supply transport, personnel movement, and equipment maintenance. AI optimizes fleet management, predicts anomalies, and improves resource allocation, generating economic savings and operational efficiency.
- Unit Training: Physical military training is costly and inherently risky. AI enables virtual and augmented reality simulations, allowing customized training based on individual combatant performance, improving readiness, and reducing risks.

### D. Organizing Criteria Answers: A Tabular Approach

The different types of responses present in the tables of this work have been classified. Now, for the explanatory section, we describe each response type and its purpose:

- Short Text: Responses of brief description, typically model names, acronyms, or general categories. This is used when a concise label sufficiently conveys the information without additional explanation.
- Descriptive Text: a more detailed explanation is needed without being too lengthy, for example, in descriptions of methodologies or advantages of particular approaches. Application and objectives descriptions or complex descriptive key factors.
- Yes/No: Used for binary responses to indicate the presence or absence of a feature. This is useful for quick verifications, such as whether a model supports a specific functionality.
- Numerical Values: Quantitative indicators such as accuracy, success rates, or performance metrics.
- N/A (Not Available): Used when information is unavailable and/or does not apply to the category. This is common in comparative tables when specific methods are not implemented across all technologies.

### V. AI-driven Tactical Communication and Networks

In this section, we revisit the military technologies and systems described in Section III, incorporating advancements and considerations where ML has been applied to specific components. These updates align with the criteria outlined in the methodology, highlighting how

ML enhances functionality and addresses challenges within these systems.

## A. Information Network Systems

As presented in Section III-A, information systems in tactical or military environments are critical tools designed to support decision-making and situational awareness, C2. These systems manage data from the rest of the systems compiled in this survey; therefore, they are fundamentals in full tactical systems.

ML can significantly enhance these systems by automating complex decision-making processes, extracting actionable insights from large datasets, and adapting to evolving threats. Some benefits of incorporating ML include:

- Enhanced Situational Awareness: ML algorithms can process sensor data to identify patterns, detect anomalies, and predict adversarial actions, improving battlefield awareness.
- Autonomous Systems: ML enables autonomous drones, surveillance systems, and robotic units to operate with minimal human intervention from information systems.
- Decision Support: ML provides commanders with data-driven recommendations by integrating predictive analytics.
- Cybersecurity: ML fortifies systems against cyber threats by detecting and mitigating unusual network behaviors.

In addition, these benefits bring transformative advantages for information systems like C2, such as i) speed, which accelerates decision-making by analyzing data in real-time; ii) accuracy, which reduces human errors in intelligence assessment; iii) scalability, which handles vast amounts of data efficiently; iv) adaptability which learns and evolves with new data to counteract emerging threats.

Various publications aim to improve information systems in military tactical environments by leveraging ML. Table II presents an analysis based on Criterion 1, highlighting the most relevant publications alongside the concepts discussed in the previous section and identifying the military areas they may impact focused on information systems. Table III provides an analysis based on Criterion 2, while Table IV focuses on Criterion 3.

RL techniques have significantly enhanced tactical communication and decision-making systems in information systems for modern warfare scenarios. RL-based algorithms optimize decentralized multi-agent communication within tactical networks, leveraging Cooperative Learning (CL) agents and tactical replay databases to manage critical metrics such as signal-to-noise ratios. In contrast, environment-dependent communication improves command support and scalability [68]. For the Internet of Battlefield Things (IoBT), ML classifiers like Support Vector Machines (SVM) and Random Forest (RF) prioritize battlefield data processing, although the absence of comprehensive military datasets limits their optimization potential [69]. Similarly, ML models address spectrum scarcity in Software Defined Radio (SDR) applications, with Naïve Bayes and Gradient Boosting enhancing spectrum detection and resource allocation, albeit with constrained performance for wide-spectrum detections [70]. In predictive systems, Artificial Neural Networks (ANN) and DL forecast enemy movements, as demonstrated by augmented reality-enabled predictive mapping of naval adversaries. DL in case of enemy naval positions, trained on game-derived datasets, highlighting applications in forecasting adversarial intentions [71]. However, scalability and broader scenario adaptability remain challenges in [71]. RF-based warfare simulations analyze armored and naval combat, optimizing resource efficiency but lacking comprehensive environmental analysis or scalability [72]. The Tactical Assault Kit-ML (TAK-ML) framework, integrating battlefield sensors with ML, facilitates real-time data harmonization and secure communication, supported by TLS/SSL configurations [73]. Intelligent aerial combat maneuvers benefit from Long Short-Term Memory (LSTM)-Deep Q-Network (DQN) models, enabling precise short-range engagements despite speed limitations in deductive decision-making [74]. Hybrid RL and probabilistic approaches in missile defense systems enhance real-time efficiency and scalability, although they face challenges related to the unpredictable nature of attacking missile trajectories [75]. Reconfigurable Intelligent Surfaces (RIS) extend tactical wireless networks, boosting spectral and energy efficiency but requiring continuous Channel State Information (CSI) for optimal functionality [76]. Hybrid AI models combining Graph Neural Network (GNN) and Deep Reinforcement Learning (DRL) improve Quality of Service (QoS) in adversarial environments, advancing routing and adversarial flow management [77]. Finally, RL-based enhancements in C2 systems automate decision-making processes and enhance operational scalability, though the lack of strategic mapping remains a limitation [78].

The studies [68]–[78] for information systems predominantly focus on ground and navy operations, with fewer addressing air and space systems as compiled in Table IV. Most references support systems for real-time command and battle information sharing. Diverse approaches, including using game theory for deception and studying spectrum manipulation, are considered for cybersecurity. Military Intelligence emphasizes satellite or UAV data to enhance predictions and integrate intelligence into tactical decisions.

Maintenance, new constructions, and logistics are less frequently covered but include emerging technologies like DT [73] for maintenance and adaptable frameworks for logistic support. Some studies suggest unique methodologies like deception tactics using false signals or prioritization models for asset defense. A few highlight potential gaps, e.g., the absence of applications in asymmetric warfare scenarios.

Blockchain technology has emerged as a promising solution for enhancing the trust, security, and decentral-ization of information systems. Its adoption is gradually being explored within the defense sector. For instance, blockchain can be leveraged to manage and coordinate information system services across federated coalition networks, enabling secure and tamper-resistant service orchestration among allied entities [79], [80], [81].

TABLE II: Criterion 1 for Information Network Systems

| Ref. | Application | Objective | Innovation | AI Type | Training data information |
|---|---|---|---|---|---|
| [68] | Decentralized multi-agent architecture to optimize communication for military applications in DIL tactical networks using CL techniques enhanced with ML. | Decentralized reinforcement-based ML approach to enhance the network, where each node is optimized by a CL agent employing RL to act based on its local observations. | Disconnected Intermittent and Limited (DIL) Networks; Command and Control Information Systems (C2IS) service layers; GNN architectures. | RL based on CL observations. | The actions performed by tactical agents, as well as the SNR ratios calculated between each pair of units and positions, are stored in the Tactical Replay database. |
| [69] | Introduction of a ML classifier to determine what type of IoBT device data to transmit on the battlefield and under what conditions. | Transforming real-time data from C4ISR IoBT devices into secure, reliable, and actionable information, as IoBT devices must exchange data and receive feedback from other devices, such as tanks and C2 infrastructure in real-time. | Command Control Communications Computers Intelligence Surveillance and Reconnaissance (C4ISR) devices, IoBT devices, JointField blockchain network. | SVM, Bayes Point Match, Boosted Decision Trees, Decision Forests, and Decision Jungles. | No specific military database was found. The study recommends conducting tests in a real-world environment. |
| [70] | Enhancing free spectrum detection using ML for SDR applications. | Providing flexibility and configurability to address spectrum scarcity in wireless communication systems. | SDR and CR networks. | Comparison of 4 supervised ML models: Native Bayes classifier, SVM, Gradient Boosting Machine, and Distributed Random Forest. | No specific military database was found. |
| [71] | Predicting enemy location in naval combat using DL. | Forecasting enemy naval positions and movements based on known locations. | Inferring adversarial intentions. | ANN and Random Forest implementations. | Models trained and tested with "World of Warships" gameplay data from former naval officers. |
| [72] | Warfare simulation to predict the winning warship using Random Forest. | Predicting the winner based on seven characteristics: size, speed, capacity, crew number, attack, additional attack, and defense. | OODA loop. | Supervised ML using Random Forest. | Using 9,660 battleship datasets (7,728 training - 1,932 testing). |
| [73] | Describing the TAK-ML framework for data collection, model building, and deployment in soldier-proximal tactical environments. | Exploiting battlefield sensor data to provide services for other applications. | Every Soldier is a Sensor (ES2), TAK ecosystems, SA. | TAK-ML harmonizes ML libraries, sensors, hardware, and applications on TAK servers. | TAK servers collect, fuse, and analyze data to enable ES2 battlefield operations. |
| [74] | LSTM-DQN algorithm and deep network for data collection, model planning issues. | Avoid enemy threats and gather advantages to threaten targets, enabling intelligent aerial combat. | BVR aerial combat; reactive and deductive decision-making. | DQN and Based on LSTM cells, where the perception layer converts basic states into high-dimensional SA. | Not specified. |
| [75] | Missile defense decision-making system in incomplete information scenarios. | Facing massive missile attacks in a short time frame. | N/A. | Hybrid method combining a prior probability hypothesis of attack and RL framework. | Recommends adding factors like missile angle and increasing missile/asset types and scales. |
| [76] | ML for RIS to enhance network capacity and coverage. | Maximizing wireless communication advantages with increased interactions. | Introducing RIS technology | SL, UL, RL, and FL. | Not specified. |
| [77] | Enhanced QoS of information operating in hostile environments that may host active adversaries | Improving QoS in tactical MANETs. | A hybrid AI model combining GNN and DRL | RNN, GNN, DRL | Not specified. |
| [78] | enhancing information in C2 systems under modern operations | automating and enhancing military decision-making in C2 systems. | decision-making with capabilities of RL. | RL | Not specified. |

### B. Image Surveillance Systems

An image surveillance system integrated with AI and ML algorithms can detect anomalies, classify targets, and predict potential risks. Thus, it is critical for reconnaissance, battlefield monitoring, and securing military installations.

Nowadays, in the commercial world, surveillance systems are installed in companies or governments that utilize ML techniques [82]. For example, there are CCTV systems where the processing and control units apply methods such as facial recognition, fingerprint identification, and automatic detection of aggressive human behavior or theft. These systems can directly request the presence of state security forces in the area, among other applications.

However, the reviewed literature contains very few specific references to defense systems. Therefore, the three most relevant studies have been selected. These studies would be highly useful in specific military environments.

ML can profoundly enhance image surveillance systems in military contexts by leveraging these systems to be able to process vast amounts of visual data, enabling:

- **Enhanced Threat Detection:** ML models, such as Convolutional Neural Networks (CNNs), can identify and classify objects like weapons, vehicles, or intruders in real-time. For instance, [83] utilized You Only Look Once (YOLO) version 2 YOLOv3 and Faster Region-based Convolutional Neural Network (RCNN) Faster-RCNN for automatic weapon detection, demonstrating the potential for rapid and accurate threat identification using CCTV feeds.
- **Improved Accuracy:** ML aids in reducing false positives and negatives by learning from diverse datasets, including thermal and infrared images. Thermal imaging applications, such as those in [84], leverage models like YOLOv8, achieving a mean Average Precision (mAP) of 96%, even in challenging envi-ronmental conditions.

TABLE III: Criterion 2 for Information Network Systems

| Ref. | Data Fusion | Tactical Scenario Inference | Command Assistance Element | Offer Roadmaps against Threats | Ergonomic Human-Machine Interface | Universal Language with Other Systems | Human Rules of Warfare | Availability | Resource Optimization | Scalability of Structures | Integrity | Reaction Speed |
|---|---|---|---|---|---|---|---|---|---|---|---|---|
| [68] | N/A | Environment-dependent communication. | Yes | Yes, decentralized architecture. | N/A | Yes, requires CL communication. | N/A | Varies by topology and link quality. | Yes, optimizes network use. | Yes, extends to nodes. | Nodes vulnerable to attacks. | Depends on nodes and CL. |
| [69] | Yes, large IoBT data. | Yes, prioritizes battlefield data. | Yes | No, lacks threat roadmap. | N/A | Yes, devices intercommunicate. | N/A | Decision Jungles are optimal. | Yes, filters massive data. | Yes, applies to scenarios. | Dynamic threat routes. | Decision Jungles optimal. |
| [70] | N/A | Yes, studies spectrum use. | Yes, finds free spectrum. | Yes, proposes zones. | N/A | Yes, 4 valid SDR algorithms. | N/A | Naïve Bayes preferred. | Yes. | Yes. | Yes. | Low for wideband. |
| [71] | Predictive map of enemy in AR. | Locates enemy. | 40% prediction with 3 games. | Only predicts position. | Overlay AR map. | N/A. | Ex-officers' decisions. | ANN superior. | ANN superior. | 6 ships, no unit expansion. | Study false signals. | ANN superior. |
| [72] | Needs 7 battleship features. | No scenario. | Missing factors. | No roadmap. | Win/loss only. | N/A | 7 features based. | Simple algorithm. | Simple algorithm. | Future work. | Too few features. | Simple algorithm. |
| [73] | Shares map, chat, video in battle. | Depends on TAK-ML app. | Image recognition. | Terrain learning. | Visual apps. | TAK-ML info-sharing. | Depends on app. | Depends on coverage. | Harmonizes data. | Harmonizes ML and apps. | TLS/SSL in TAK-ML. | Depends on app and coverage. |
| [74] | Used in flight and motion models. | Basic combat model: flight, motion, missile. | Assists pilot. | Evades threats, guarantees position. | Graphical. | Automatic alternative. | Motion models, missile envelopes. | Short-range combat focus. | N/A | N/A | Calculates best tactics. | Slow if using deductive decisions. |
| [75] | Merge attack, defense, and missile layers. | Unknown missile distribution. | Attack alternatives. | Optimizes defense missile allocation. | Graphical. | Integrates defense system. | Needs asset list. | Hybrid method surpasses heuristic methods, DQN. | Uses only necessary defense. | Adapts to available missiles. | Low, missile distribution unknown. | Hybrid method enables real-time deployment. |
| [76] | RIS retransmits data. | Poor CSI must be addressed. | Extends battlefield wireless network. | N/A | N/A | Interacts with other systems. | N/A | Needs accurate channel info. | Improves energy efficiency. | Scalable RIS structure. | Enhances link quality. | Needs real-time data. |
| [77] | Adversary flow data. | Active adversary data. | No specification. | Improves defense. | Graphical implied. | Must communicate with C2. | N/A | Not specified. | QoS optimization. | No | Predicted by network. | Slow info update. |
| [78] | From various operations. | Central C2 system. | Through C2. | Should consider it as C2. | C2 system. | Not specified. | N/A | Not specified. | No | Yes | Very robust. | Fast, real-time. |

TABLE IV: Criterion 3 for Information Network Systems

| Ref. | Army | War Strategies | Command Support | Cybersecurity | Military Intelligence | New Constructions | Air Operations | Ground Operations | Naval Operations | Logistics | Unit Training |
|---|---|---|---|---|---|---|---|---|---|---|---|
| [68] | Land | Yes | Yes | Study node/CL manipulation | Yes | N/A | Possible, few nodes | Yes | Possible, few nodes | Yes | Yes |
| [69] | Land | Yes | Yes | Use game theory for data deception | Yes | N/A | N/A | Yes | N/A | Yes | Yes |
| [70] | Land | N/A | Yes, free spectrum system | Study spectrum manipulation | Yes, free spectrum system | N/A | Possible | Yes | Possible | Possible with fewer devices | Possible |
| [71] | Navy | Yes | Yes | False signals for deception | Prediction with satellite/UAVs | New naval platforms incentive | Algorithm modifiable | Algorithm modifiable | Yes | Yes | Yes |
| [72] | Navy | Yes, Random Forest 80% accuracy | Yes | N/A | Yes | New naval platforms incentive | Algorithm modifiable | Algorithm modifiable | Yes | Yes | Yes |
| [73] | Land | Yes, depends on application | Yes, depends on application | Study false signals or TLS/SSL enabled | Yes, depends on application | Yes, new applications possible | No, primarily | Yes | Yes, for asymmetric warfare | Yes, depends on application | Yes, depends on application |
| [74] | Air and Space | Yes, aerial strategy for missile launch | Yes, action plan for aircraft | Study interference in decision-making | N/A | N/A | Yes | Algorithm modifiable | Algorithm modifiable | N/A | Yes, adapts to simulations |
| [75] | Land | Yes, prioritize assets | Yes, missile allocation | N/A | Converge with military intelligence | N/A | Yes | Algorithm modifiable | Algorithm modifiable | N/A | Yes, adapts to simulations |
| [76] | Land | Yes, wireless network for battlefield info | Yes, real-time battle info | Study wireless interference | Yes, RIS with UAVs | Yes | Yes | Yes | Yes | N/A | Yes, train network usage for battle |
| [77] | Land | No | Yes | No | Yes | N/A | Yes | Yes | Yes | N/A | N/A |
| [78] | All | Yes | Yes | No | No | N/A | Yes | Yes | Yes | No | No |

- Anomaly Detection: Algorithms like AutoEncoder (AE) can identify unusual activities or objects, enhancing perimeter security. This capability has been effectively demonstrated in radar-based applications [85], where noise-removal AE improved underwater image quality for better anomaly detection.
- Operational Efficiency: Autonomous systems, powered by ML, can monitor areas continuously with minimal human intervention, optimizing resource utilization. For example, the drone detection systems in [86] employed Faster-RCNN and YOLOv3 models to enable high-accuracy UAV tracking in diverse aerial scenarios.
- Data Fusion and Tactical Insights: ML enables the fusion of multimodal data, such as infrared and visible images, to provide more precise and more informative surveillance outputs. As shown in [87], Deep Supervised Generative (DSG)-Fusion techniques allow for integrating multiple image sources, aiding in tactical scenario analysis.
- Resource Optimization and Scalability: These systems can scale efficiently to handle increasing data loads while maintaining high performance. For instance, [88] demonstrated real-time military aircraft detection using TensorFlow-based CNNs on large annotated datasets, ensuring scalable and reliable surveillance operations.

These advantages make ML indispensable for modernizing military surveillance and addressing challenges such as diverse terrains, environmental conditions, and evolving threats. Adopting novel ML architectures, including Generative Adversarial Networks (GANs) and advanced pre-processing techniques, has enabled the development of robust, efficient, and scalable systems tailored to specific military needs. Additionally, the integration of ergonomic human-machine interfaces [86] and real-time alert systems [83] underscores the potential for ML-driven solutions to transform military surveillance capabilities comprehensively.

Various publications aim to improve image surveillance systems by leveraging ML. Table V presents an analysis based on Criterion 1, highlighting the most relevant publications alongside the concepts discussed in the previous section and identifying the military areas they may impact. Table VI provides an analysis based on Criterion 2, while Table VII focuses on Criterion 3.

### C. Electronic Warfare Systems

ML has a wide field of applications in EW. Within the ESM field, it could automate the identification and search for radioelectric emissions by studying and learning the available libraries. In the ECM area, it could recommend the type of countermeasure the strategy to be used and, finally, related to EPM, based on the disturbance received, it could automate the frequency hopping of the agile frequency transmitting radars to avoid being canceled.

Given that, incorporating ML in EW Systems offers several significant advantages that can be summarized as follows:

- Enhanced Threat Detection: ML algorithms improve threat detection and classification by analyzing vast datasets to identify patterns and anomalies that may indicate hostile activities. This capability allows for real-time adaptive responses, improving the system's effectiveness in dynamic environments.
- Automation: ML facilitates the automation of signal processing tasks, reducing the cognitive load on human operators and increasing operational efficiency. By learning from historical data, ML models can predict and counteract enemy tactics, providing a strategic advantage.
- Adaptation: ML-driven EW systems can continuously evolve, adapting to new threats and technologies without requiring extensive reprogramming. This adaptability ensures that military forces maintain a technological edge over adversaries.
- Integration: The integration of ML into EW systems supports the development of more sophisticated jamming and deception techniques, enhancing the ability to disrupt enemy communications and radar systems. ML significantly increases EW operations' capability, adaptability, and resilience.

Thus, different works in the literature aim to improve EW by leveraging ML. Table VIII presents an analysis based on Criterion 1, highlighting the most relevant works alongside the concepts discussed in Section IV and identifying the military areas they may impact focused on EW. Table IX provides an analysis based on Criterion 2, while Table X focuses on Criterion 3. Note that some of the publications presented show an N/A in some criteria due to their lack of relevance.

These publications collectively underscore the advancements in EW systems by integrating ML, improving decision-making, threat detection, and operational efficiency. An integrated ML-assisted EW system that autonomously navigates and assesses threats using cognitive and multimode radar systems is discussed in [91]. In [92] is presented a 3D Explorer Space Program for simulating Cognitive Electronic Warfare (CEW) environments with UAVs, focusing on threat detection and countermeasure selection using DRL algorithms. [93] addresses the construction of reduced electromagnetic wave shape models to improve computational efficiency in radar and EW simulations. A CNN-based method is presented in [94] for classifying radar interference signals, emphasizing using Siamese-CNN for effective classification with limited training samples. [95] explores ML-based Global Navigation Satellite System (GNSS) models to enhance signal robustness and performance in hostile environments, utilizing various ML techniques. Deep learning methods are detailed in [96] for predicting interference techniques, employing DNN and LSTM networks for accurate threat response. CNN-based radio fingerprinting is focused in [97]

TABLE V: Criterion 1 for Image Surveillance Systems

| Ref. | Application | Objective | Innovation | AI Type | Training data information |
|---|---|---|---|---|---|
| [83] | Automatic weapon detection in real-time using CCTV videos. | Balances real-time performance with accuracy. | Region of Interest (ROI)-based object detection. | VGG16, YOLOv3, YOLOv4, etc. | Custom dataset from web and videos. |
| [85] | Improving laser-coded images for underwater systems. | Resolves turbid environment challenges. | Noise removal autoencoder. | Shallow and deep networks. | Lab-collected images. |
| [87] | Fusion of infrared and visible spectrum images. | Texture retention in fused images. | Double-flow guided filter. | VGG, GAN. | Not specified. |
| [86] | Autonomous drone detection and tracking. | High accuracy with optimized memory. | Unified Object Scale Optimization. | YOLOv3, Mask R-CNN. | Kaggle/custom drone images. |
| [84] | Thermal human detection for security operations. | Achieves 96% mAP. | YOLOv8 for thermal imaging. | YOLOv7, YOLOv8. | Augmented thermal datasets. |
| [89] | Foreign object detection in radio imaging. | Accurate in noisy environments. | YOLOv3 applied to radar images. | CNN, YOLOv3. | Synthetic RF radar images. |
| [90] | Military vehicle recognition using small datasets. | High resource demand for neural networks. | Transfer learning with ResNet50. | ResNet50, Xception. | Social media images with augmentation. |
| [88] | Military aircraft detection for real-time surveillance. | Reliable under varying conditions. | TensorFlow-based pre-processing. | CNN. | Large dataset with annotations. |

TABLE VI: Criterion 2 for Image Surveillance Systems

| Ref. | Data Fusion | Tactical Scenario Inference | Command Assistance Element | Offer Roadmaps Against Threats | Ergonomic Human-Machine Interface | Universal Language with Other Systems | Human Rules of Warfare | Availability | Resource Optimization | Scalability of Structures | Integrity | Reaction Speed |
|---|---|---|---|---|---|---|---|---|---|---|---|---|
| [83] | Multi-angle images | Weapon occlusion issues | Defense system applications | Alerts in case of threat | Not indicated, but should be ergonomic | Associated with lighting/weapons | N/A | Real-time | N/A | More classes needed | 99% confidence score | Improve precision |
| [85] | Laser images | Better scattering suppression | Enhances underwater imaging | N/A | Ergonomic and visual interface | Sonar compatibility | N/A | Improves image quality by 25% | Extends detection range in turbid zones | N/A | No behavior under disturbance | Not indicated |
| [87] | Infrared and visible fusion | Influenced by atmospheric conditions | Aids in target identification | N/A | Good visual interface | Fire control/CCTV application | N/A | Real-time | N/A | Scales to CCTV systems | No countermeasures for heat deception | Low latency |
| [86] | Multi-frame fusion for drones | Effective in aerial scenarios | Enhances real-time drone detection | Warns UAV threats | User-friendly platform | Compatible with multi-sensor tracking | Designed for military compliance | High frame rate | Memory optimization | Tracks multiple drones | High UAV detection reliability | Low-latency detection |
| [84] | Thermal imaging datasets | Robust in fog/smoke | Improves human detection | Enhances threat awareness | Interpretable thermal imagery | Security infrastructure integration | N/A | Real-time with YOLOv8 | Processing efficiency | Supports larger datasets | 96% mAP | Real-time alerts |
| [89] | Radar imaging | Reliable in noisy environments | Detects concealed objects | Identifies explosives | Intuitive radar interface | Adaptable for radar warnings | N/A | Real-time imaging | Minimal dataset reliability | Various radar applications | Reliable detection | Quick response |
| [90] | Augmented datasets | Identifies vehicles in dynamic settings | Aids tactical decisions | Vehicle threat insights | Enhances situational awareness | Integrates with vehicle systems | Standards compliance | Real-time processing | Efficient dataset use | Scales to fleets | High accuracy in conditions | Minimal latency |
| [88] | Diverse perspectives | Aircraft detection in varied conditions | Precise localization | Aerial threat alerts | Simplifies detection | Coordinates with monitoring systems | Standards compliance | Rapid detection | Resource-efficient | Large-scale monitoring | Reliable in challenges | Fast processing |

TABLE VII: Criterion 3 for Image Surveillance Systems

| Ref. | Army | War Strategies | Command Support | Cybersecurity | Military Intelligence | New Constructions | Air Operations | Ground Operations | Naval Operations | Logistics | Unit Training |
|---|---|---|---|---|---|---|---|---|---|---|---|
| [83] | Army | CCTV detects armed personnel. | Confirms threats using data. | CCTV dependent systems. | Aids monitoring in zones. | N/A | N/A | N/A | N/A | N/A | Systems trained for monitoring. |
| [85] | Army | Supports command weapon data. | Improves tactical decisions. | N/A | Captures in turbid conditions. | N/A | N/A | N/A | N/A | N/A | Training for imaging systems. |
| [87] | Army | Captures clearer images. | Better tactical insights. | N/A | Alternate viewpoints support missions. | N/A | N/A | N/A | N/A | N/A | Operational training needed. |
| [86] | Army | Tracks UAVs in combat. | Drone data aids command. | N/A | Predicts aerial threats. | Improves UAV usage in combat. | N/A | N/A | N/A | N/A | Drone training readiness. |
| [84] | Army | Thermal systems detect threats. | Alerts for low visibility risks. | N/A | Identifies targets in fog. | Rescue aid in missions. | N/A | N/A | N/A | N/A | Thermal training improves. |
| [89] | Army | Radar images find IEDs. | Supports precise commands. | Enhances detection. | Tracks threats with accuracy. | Urban detection tools aid ops. | N/A | N/A | N/A | N/A | Improves system training. |
| [90] | Army | Vehicle ID in combat. | Better fleet decisions. | Secure data usage. | Military vehicle insights. | Scalable tracking systems. | N/A | N/A | Logistics aid fleet tracking. | Fleet-based training. |
| [88] | Army | Aircraft tracking aids ops. | Precise aerial surveillance. | Relies on secure channels. | Supports tactical analysis. | Insights for aerial ops. | N/A | N/A | N/A | Aerial imagery training. |

TABLE VIII: Criterion 1 for Electronic Warfare Systems

| Ref. | Application | Objective | Innovation | AI Type | Training data information |
|---|---|---|---|---|---|
| [91] | Integrated ML-assisted system. | Detect and combat hostile radars, ML to classify signals, CEW system. | Cognitive and multimode radar systems. | Automatic decision tree generator, diffuse logic model and LSTM. | Decision tree is automatically generated from simulated EW encounters and data. |
| [92] | 3D Explorer Space Program to Simulate CEW Environments. | Stand-alone threat detection decision process, classification and countermeasure selection. | CEW tasks with DRL Algorithm. | Variational Bayesian method. | Deep Deterministic Policy Gradient Algorithm (DDPG). |
| [93] | Reduced electromagnetic wave shape model construction, for radar and EW simulations. | Improve computational efficiency of radar and EW simulations. | Coupling between representation and algorithms operating on representation. | Supervised and Unsupervised Machine Learning. | It is not contemplated. Underlying sources of error were identified. |
| [94] | CNN-based method for classifying radar interference signal. | 1D-CNN designed to classify radar interference signals. | Siamese-CNN (S-CNN) | SVM, Decision Tree Classifiers, Logistic Regression and RF. | Limited training samples. A CNN-based simesan network. |
| [95] | ML-based GNSS models to improve robustness and position signal performance. | Deliver low-cost, high-performance solution. | Positioning, Navigation and Timing, Time to First Correction (TTFF) | 213 application studies from 2000 to 2021. Mostly used RF, SVM, ANN and CNN. | |
| [96] | Two deep learning-based methods for predicting the proper interference technique. | DNN on manually extracted feature values from the PDW list and using LSTM that takes the PDW list as input. | Press Description Word (PDW), Long Short-Term Memory (LSTM). | DNN of different structures and LSTM. | Training data built from the library. Trained ML model to predict interference technique. |
| [97] | CNN-based radio fingerprinting. | Timely interception of tactical and strategic transmissions. | Transform the identification and classification of RF signals. | CNN. | IQ Data and Image Processing. |

to intercept crucial transmissions and segregate radios based on significance, enhancing radio-frequency signal identification and classification.

The technologies used for EW systems integrate with or operate in conjunction with the other systems explained here. They are mainly combined with the radar system discussed in the following section.

### D. Radar Systems

Radar systems have undergone significant advancements in recent years. Authors in [98] provide a comprehensive overview of AI approaches to enhance radar data processing tasks. These approaches can refine existing methods or even replace conventional techniques with more powerful alternatives. For instance, this study [98] explores methods for identifying disruptions in air traffic control, distinguishing legitimate targets from parasitic echoes such as weather phenomena or bird activity. Additionally, it delves into marine environments, focusing on differentiating land clutter, calm sea clutter, rough sea clutter, and composite clutter.

From a radar signal processing perspective, the authors in [99] analyze the role of ML in military services, among other applications. CNN and SVM are the primary techniques suggested for improving radar signal processing in these contexts.

Integrating ML into radar systems is not merely a technological advancement but a necessity to meet the challenges posed by modern operational environments. Traditional radar techniques, while effective, are often limited in handling the increasing complexity of tasks such as:

- Clutter Suppression: Conventional algorithms struggle to differentiate meaningful targets from environmental noise or clutter in dynamic scenarios like urban areas or rough seas. ML models, trained on diverse datasets, excel at recognizing patterns and suppressing noise, thus improving target detection accuracy.
- Real-Time Adaptability: Radar systems must adapt to rapidly changing environments, such as varying

weather conditions or evolving combat scenarios. ML enables systems to learn and adjust quickly, enhancing situational awareness and decision-making capabilities.
- Automation of Complex Tasks: Modern radar systems handle large volumes of data, requiring efficient automation of tasks like anomaly detection, predictive maintenance, and data fusion. ML algorithms provide the computational power and intelligence to automate these processes without compromising accuracy.

We can find the following benefits of ML for Radar Systems:

- Robust Performance in Complex Environments: ML models can adapt to complex scenarios, including multi-path effects, electromagnetic interference, or high-clutter environments, maintaining high performance where traditional methods falter.
- Enhanced Detection and Classification: ML algorithms significantly improve the detection and classification of targets by learning from extensive datasets. This is particularly useful in distinguishing between similar objects, such as UAVs and birds, or identifying subtle changes in terrain.
- Predictive and Proactive Capabilities: Incorporating ML allows radar systems to predict potential issues, such as equipment failures or evolving threats, enabling proactive measures.
- Increased Efficiency: By automating repetitive or computationally intensive tasks, ML reduces the workload on human operators and accelerates processing speeds, making real-time analysis feasible.

In the field of radar systems, there is extensive literature on applications outside the defense domain. However, its application in this context is more limited. For this reason, we have focused in this work on its specific application in tactical environments and conducted a detailed study. Following the established criteria in Section IV for analyzing ML applications in defense, the analysis for radar systems is summarized in Table XI, Table XII, and Table XIII, corresponding to Criteria 1, Criteria 2, and Criteria 3, respectively.

TABLE IX: Criterion 2 for Electronic Warfare Systems

| Ref. | Data Fusion | Tactical Scenario Inference | Command Assistance Element | Offer Roadmaps against Threats | Ergonomic Human-Machine Interface | Universal Language with Other Systems | Human Rules of Warfare | Availability | Resource Optimization | Scalability of Structures | Integrity | Reaction Speed |
|---|---|---|---|---|---|---|---|---|---|---|---|---|
| [91] | Yes, environments difficult to analyze. | Yes, systems must accurately model environment. | Yes, system detects, acquires and follows goals, and guides the platform. | Yes, the Electronic Attack (EA) assessment model recommendations. | The interface is not specified by supposedly visual. | Yes, with other radar systems and countermeasures. | Observed radar threat level determined by distance and mode. | Satisfactory results in simulated environment against multiple multifunction radars. | Better result with short-term memory neural network. | Yes, focus allows automatic updates in progress. | Each case was simulated 100 times and successful missions were recorded. | Depends on the distance at which the location of the threat is assumed. |
| [92] | UAV speed and direction of motion need to be controlled while verifying the integrity of the CEW system function. | You can search for the station in the mission map + Bayesian inference or applying evolutionary computing method. | Yes, to locate ground stations that are transmitting. | No, it is not contemplated. | No, it is not considered. | Both the UAV and ground station use radar-like observation sensors. | In this case, it is a game-like simulation. | Partial observability of the environment, and physical UAV maneuverability restrictions. | No, it is not considered. | Detection sensor modules and countermeasure weapons can be expanded. | The interaction between each part and the environment has a clear mathematical model. | UAV movement prioritizes physical restrictions of movement over CEW system operation. |
| [93] | Yes, there is a representative coupling of the sampled signal and morphism. | N/A | N/A | N/A | N/A | Seeks to formulate learning problem in a unified way to increase efficiency. | N/A | Loss of information due to morphism and error in the approximations of the supervised learning algorithm. | Yes, morphism avoids computational bottleneck. | Learning problem should be formulated in a unified way to increase the effectiveness of the outcome. | Worsens by being more truthful about sampled representations than morphism. | Improvement with approximate morphism based on features of a reduced model. |
| [94] | S-CNN to classify different interference signals with limited samples. | N/A | Yes, this electromagnetic signal classification method can give enemy information. | N/A | N/A | N/A | N/A | Yes, 1D-CNN experimental result. S-CNN result under limited training samples. | N/A | N/A | 12 typical types of radar interference signals. | N/A |
| [95] | N/A | Yes, GNSS in both indoor and outdoor environments. | Yes, early detection of faults and errors can lead to timely correct it. | N/A | N/A | N/A | N/A | Reduce maintenance effort and downtime. | Yes, it is the goal with ML. | N/A | Sources of errors exist for satellite-based positioning. | Dependent, SVM speed does not meet the real-time requirements for interference monitoring. |
| [96] | Yes, ML model generates interference techniques for incoming threat signals. | N/A | Yes, predicts proper interference technique in the face of a threat. | Yes, suggest an interference technique before a threat. | N/A | N/A | N/A | Interference method can be predicted for unknown radar signal with an average accuracy of about 92%. | Yes, the predicted interference method will be used first. | N/A | Prediction accuracy of the LSTM model was higher. | DNN-based method is faster than LSTM method. |
| [97] | IQ Data and Image. | Hybrid multi-level approach for fingerprinting and confirming transmitter identification. | Yes, confirming transmitter identification in dynamic and wider spectrum. | Yes, identifying high-value targets and assisting in Identification Friend or Foe, anti-spoofing. | N/A | N/A | N/A | Presented concepts and solutions can be a game changer for both military and civilian use. | Yes, the hybrid approach reduces computational load. | N/A | Consistency in accuracy in SNR levels and BW. IQ and Image processing-based models showed an exponential decline in accuracy. | N/A |

In [99], condition-based maintenance for air defense radar systems is explored utilizing a variety of ML models like RF, Multi-Layer Perceptron (MLP), and XGBoost to distinguish between malfunctions and normal conditions. Another reference, [100], highlights the use of CNNs in Synthetic Aperture Radar (SAR) for automatic target recognition and classification of land types. Similarly, [101] uses CNNs in radar resource management to enhance performance under high-target loads, while [102] focuses on improving the detection and classification of airborne targets using CNNs for real-time air traffic control. Other references also employ ML techniques, such as unsupervised learning and CNN-based detectors, to address issues like computational complexity, noise reduction, and radar accuracy in various operational scenarios. [104] focuses on developing unimodular waveforms for Multiple-Input Multiple-Output (MIMO) radar to enhance localization accuracy, clutter mitigation, and Doppler ambiguity reduction. It employs a deep residual network-based optimization approach and uses the Adam algorithm for unsupervised optimization. A CNN-based detector called RadCNN is introduced in [105], replacing standard Constant False Alarm Rate (CFAR) detectors in pulsed Doppler radar. RadCNN improves performance in low Signal-Noise Ratio (SNR) scenarios with significantly reduced computational complexity, leveraging 182,000 training samples for evaluation. Lastly, [106] discusses SAR in fighter aircraft by reducing the time complexity in processing radar cross-section matrices through optimized clustering techniques, utilizing methods like K-Means and Ellipsoidal Radar Cross Section (RCS) modeling to classify data into nine clusters.

Key highlights in Table XII include data fusion capa-bilities, with several approaches integrating multiple data streams and leveraging advanced ML methods, such as CNNs for SAR domains [100] and MIMO radars [101]. Tactical scenario inference is unevenly addressed; while ML aids in overloading scenarios [101], others focus on improving detection accuracy [105]. Command assistance is emphasized in systems improving target classification [102] or enabling real-time responses [105]. Resource optimization and scalability vary significantly, with some approaches emphasizing low computational complexity [105] or adaptive algorithms for efficiency [103]. Integrity and reaction speed are enhanced in systems using noise reduction techniques and high-performance ML algorithms [102], [104]. Compatibility with other systems, ergonomics, and adherence to human warfare rules are less consistently addressed, highlighting lines for future advancements.

TABLE X: Criterion 3 for Electronic Warfare Systems

| Ref. | Army | War Strategies in Armed Conflicts | Command Decision Support | Cybersecurity | Military Intelligence | New Constructions | Air Operations | Ground Operations | Naval Operations | Logistics | Unit Training |
|---|---|---|---|---|---|---|---|---|---|---|---|
| [91] | Air and Space. | Yes, electromagnetic signal capture and classification is vital. | Yes, an imminent danger detected anticipates decisions. | N/A | Yes, although the algorithm collects real-time data there is a signal database. | Yes, new UAV capability and drives projects like NGJ-MB. | Yes | Could be applied. | Could be applied. | N/A | Yes, operator training would be convenient. |
| [92] | Air and Space. | Yes, ground station transmission detection-interference is vital. | Yes, an imminent danger detected anticipates decisions. | N/A | Yes, location of transmitting earth stations. | Yes, new UAV capacity. | Yes | Could be applied. | Could be applied. | N/A | Yes, operator training would be convenient. |
| [93] | Air and Space. | N/A | N/A | N/A | N/A | Yes, it could be considered in new radar and EW processing. | May be applied on radar and EW. | May be applied on radar and EW. | May be applied on radar and EW. | N/A | N/A |
| [94] | Ground | Yes, signal classification gives enemy information. | Yes | N/A | Yes, signal classification gives enemy information. | N/A | Yes | Yes | Yes | N/A | N/A |
| [95] | Ground | Yes, may cause interference with GPS signal from units. | Yes, an accurate signal is needed. | N/A | Yes, identifies unit positions. | N/A | Yes | Yes | Yes | N/A | N/A |
| [96] | Ground | Yes, can cause radar interference or enemy EW. | Yes, it prepares interference system such as a pitcher. | N/A | Yes, it could be associated with the military and with the interference system. | N/A | Yes | Yes | Yes | Yes, you can predict which interference systems to have armed. | Yes, its use should be trained. |
| [97] | Air, Ground and Space. | Yes, interception of crucial tactical and strategic transmissions. | Yes, RF fingerprinting serves as a cornerstone in ensuring seamless operations. | Security and operational integrity by RF fingerprinting. | Yes, a comprehensive solution for fingerprinting and confirming transmitter identification. | Yes, considering the merits and drawbacks of previous approaches, a hybrid approach is proposed. | Yes | Yes | Yes | Yes, RF fingerprinting in fortifying security and operational integrity within the EW spectrum. | N/A |

TABLE XI: Criterion 1 for Radar Systems

| Ref. | Application | Objective | Innovation | AI Type | Training Data |
|---|---|---|---|---|---|
| [99] | Maintenance of air defense radar systems | Detect faults | Fault detection using multiple ML models | Random Forest, MLP, etc. | Data collected in faulty and normal states |
| [100] | SAR data analysis for HD imaging | Automatic target recognition | CNN for noise removal and segmentation | CNN, RNN, AE, etc. | Data augmentation for SAR-ATR |
| [101] | Radar resource management | Improve reaction time and integrity | Multifunction radar with adaptive features | CNN | Limited data from exceptional cases |
| [102] | Airborne target detection | Classify targets as fixed- or rotary-wing aircraft | Two-stage CNN for noise filtering | CNN | 83,740 Doppler images |
| [103] | Radar scan clustering | Reduce dataset size | Density-based clustering algorithm | Unsupervised learning | Real dataset |
| [104] | MIMO radar waveform design | Improve localization and clutter mitigation | Deep residual network optimization | DL (CON model) | Adam algorithm with unsupervised Cost Objective Function (COF) |
| [105] | Pulsed Doppler radar detection | Improve CFAR performance | RadCNN for low SNR scenarios | CNN | 182,000 files for training/testing |
| [106] | Situational awareness | Reduce radar cross-section processing time | Optimized ML-based clustering | K-Means | 9 homogeneous and heterogeneous clusters |

Table XIII summarizes radar systems' contributions to war strategies, decision support, and operational activities. Notable findings include the use of radar in air and space operations, often integrating advanced image processing, terrain analysis, and target identification techniques for command support and intelligence [100], [101], [102]. Some studies emphasize their ability to provide real-time data for secure decision-making [105], while others highlight specific adaptations for combat, such as handling rotary-wing threats [102]. Applications extend to logistics, unit training, and maintenance, although cybersecurity and new construction coverage is limited. Emerging research points to training requirements for algorithm use and deception strategies [103], [104]. These systems enhance battlefield communication and precision, aiding pilots and operators in dynamic scenarios [99], [106].

TABLE XII: Criterion 2 for Radar Systems

| Ref. | Data Fusion | Tactical Scenario Inference | Command Assistance Element | Offer Roadmaps against Threats | Ergonomic Human-Machine Interface | Universal Language with Other Systems | Human Rules of Warfare | Availability | Resource Optimization | Scalability of Structures | Integrity | Reaction Speed |
|---|---|---|---|---|---|---|---|---|---|---|---|---|
| [99] | Yes, streams from different states | No | No | No | No | Yes, with EW | N/A | Medium | No | Future works | True positive rate: 0.84 | N/A |
| [100] | Feasibility of transferring CNN learning to SAR | No, potential improvement | Yes, visual info for command | Continuation of development | N/A | Neural networks compatible with systems | N/A | Uses large-scale datasets | Feasibility of CNN learning in SAR | Video application possible | Testing of DL algorithms needed | N/A |
| [101] | Yes, MIMO radars | Yes, ML for overloaded radars | Yes, exploration system for ESM, EA, communication | N/A | N/A | Yes, integrates ESM/EA | Yes, EA-specific | Low, no physical testing | Radar has multiple functions | N/A | Yes, ML improves integrity | N/A |
| [102] | Two-dimensional radar, single/series pulses | N/A | Yes, identifies aircraft type | N/A | Two-stage noise removal | N/A | N/A | Higher, reduces filtering stages | Eliminates CFAR/peak detection | N/A | 96% target hits, 85% noise | Higher, discards 30% noise |
| [103] | PDF projects dataset into 1D space | Adaptive cluster extraction | Yes, dense target info extraction | N/A | N/A | N/A | N/A | Higher, reduces time | Determines parameters for clustering | N/A | Efficient density-based scanning | Fast adaptive mean-shift algorithm |
| [104] | MIMO radar waveform facilitates implementation | N/A | Yes, improved waveform adds capabilities | N/A | N/A | Yes, integrates communication systems | N/A | Superior performance, acceptable optimization time | Solves waveform design problem | Possible use in warfare | Waveform enhances integrity | Feasible within computational complexity |
| [105] | N/A | Yes, affects detection results | Yes, real-time response | N/A | Real-time noise filtering | N/A | N/A | Outperforms CFAR techniques | RadCNN reduces complexity | N/A | RadCNN superior to state-of-art | Real-time feasibility |
| [106] | Single data stream | Full SA | No | Yes | No | No | Full | No | No | No | Residual error: 118807.63 | N/A |

## E. Fire and Weapon Direction Systems

Fire and weapon control or direction systems are critical technologies in tactical environments designed to enhance precision, efficiency, and situational awareness (SA) during combat operations. Integrating ML into these systems represents a transformative leap, enabling advanced capabilities such as real-time target recognition, predictive analytics, and adaptive decision-making. ML significantly enhances the flexibility and adaptability of fire and weapon control systems, providing a competitive edge over traditional setups by automating decision processes and reducing human error. This shift ultimately leads to more effective tactical responses and improved mission outcomes.

Advancements in aiming and fire systems in tactical environments through ML include:

- Ballistic trajectory prediction: DNN forecast trajectories, considering variables like wind, temperature, and material resistance (e.g., Conditional Generative Adversarial Networks (cGANs), for ballistic materials).
- Weapon and target recognition: CNN-SVM models classify and prioritize targets in real time for tactical optimization.
- Optimal firing solutions: Genetic algorithms (GA), particle swarm optimization (PSO), in an improved version Levy Flight Particle Swarm Optimization (LFPSO), combined GA-LFPSO enable precise firing configurations.
- Predictive maintenance: LSTM and Recurrent Neural Networks (RNNs) detect weapon system faults, enhancing reliability.

These innovations improve precision, energy efficiency, and adaptability in tactical scenarios.

Recent studies indicate that DL models can improve detection and response times in dynamic scenarios, significantly enhancing system reliability and reducing human error [113]. However, challenges include the need for extensive training datasets, potential adversarial vulnerabilities, and computational overhead. Unlike traditional systems, ML-powered solutions provide greater adaptability and efficiency in evolving battlefield conditions [113].

According to criterion 1, Table XIV highlights distinct applications and techniques for fire and weapon systems such as [107] employs cGANs for ballistic material prediction, using 50,000 iterations across ballistic classes; [108] predicts optical power for Free Space Optics (FSO) in maritime settings, testing K-Nearest Neighbors (KNN), RF, and ANN with year-long environmental data. [109] uses LSTM for fire system health monitoring. Combat efficiency with gyroscope data-based GOA-RNN fault prediction is introduced as an innovation in [110]. GA-LFPSO for artillery firing solutions is adopted in [111]. [112] evaluates CNN-SVM for ML-driven weapon recognition across five weapon categories. Each innovation offers operational accuracy and computational efficiency while reflecting varied ML advancements in tactical environments.

TABLE XIII: Criterion 3 for Radar Systems

| Ref. | Army | War Strategies in Armed Conflicts | Command Decision Support | Cybersecurity | Military Intelligence | New Constructions | Air Operations | Ground Operations | Naval Operations | Logistics | Unit Training |
|---|---|---|---|---|---|---|---|---|---|---|---|
| [99] | Air | No specified | Yes, evaluation of an air defense system | Not included | No | No | Yes | No | No | N/A | N/A |
| [100] | Air and Space | Yes, they depend on the images. | Yes, decisions based on images. | It will depend on the location of the data processor. | Yes, image segmentation, change analysis, terrain analysis. | N/A | Yes | Yes | Yes | N/A | Yes, its use should be trained. |
| [101] | Air and Space | Yes, ESM and EA. | Yes, exploration and communication. | N/A | Yes, database. | N/A | Yes | Yes | Yes | N/A | Yes, its use should be trained. |
| [102] | Air and Space | Yes, the presence of many rotary-wing enemies (including UAVs) changes the command strategy. | Yes, different ways of attacking fixed-wing to rotary-wing targets. | N/A. | Yes, identifies the type of aircraft. | Yes, implies new developments to avoid such classification. | Yes | Yes | Yes | N/A | Yes, its use should be trained. |
| [103] | Air and Space | N/A | Yes, identifies targets in dense areas. | N/A | N/A | N/A | Yes | Yes | Yes | N/A | Yes, the way to deceive the algorithm could be trained. |
| [104] | Air and Space | Yes, it can provide communication in battle and greater precision. | Yes, greater security in decision-making. | N/A | Yes, more accurate information. | N/A | Yes | Yes | Yes | N/A | Yes, the way to deceive the algorithm could be trained. |
| [105] | Air and Space | N/A | Yes, real-time information. | N/A | Yes, more information available with less SNR ratio. | N/A | Yes | Yes | Yes | N/A | Yes, the way to deceive the algorithm could be trained. |
| [106] | Air | Pilots to assess, anticipate, and respond adeptly to dynamic combat scenarios. | Help to Pilots | No | No | No | Yes | No | No | N/A | N/A |

TABLE XIV: Criterion 1 for Fire and Weapon Direction Systems

| Fire and Weapon Direction System | | | | | |
|---|---|---|---|---|---|
| Ref. | Aplication | Objective | Innovation | AI Type | Training data information |
| [107] | Successfully predict the dynamic response of materials to ballistic impacts. | Characterize material behavior across a range of loading rates and impact scenarios. | Ballistic Limit Velocity (BLV). | Conditional Generative Networks (cGAN). | Trained 5 GAN models with 50,000 iterations. Each model was trained in a multi-class format with 10 classes of ballistic data labeled from 0 to 9. |
| [108] | Accurate prediction of optical power for an FSO link in a maritime environment. | Use optical power prediction algorithms to calculate the optical power required for a high-energy laser weapon. | Power required for a laser weapon. | Studied 5 algorithms: KNN, Decision Tree, Gradient Boosting Regression, RF, and ANN. | Training data obtained over 12 months from a commercial FSO system and a weather station. |
| [109] | Fault Detection | Requirements to integrate AI | Improving the health of fire systems | LSTM,DL | Not specified |
| [110] | Fault Prediction | Improving combat efficient in various vehicles | Introducing the prediction in control system of armoured vehicles | GOA-RNN | database of data from gyroscope, using 70% for training |
| [111] | Artillery Firing Data Solving Method | Speed up convergence, robust search capabilities in solving the firing data | Improving upon issues such as population initialization, local optimum problems, and calculation efficiency. | GA-LFPSO | Ballistic model |
| [112] | AI-Driven Weapon Recognition | Improving accuracy, performance, scalability, and resistance to environmental variability | Evaluating a dual-framework methodology to assess the effectiveness of different ML techniques in detecting weapons from five unique categories | CNN-SVM | dataset includes five categories of weapons: firearms, edged weapons, explosives, improvised types of weapons, and chemical ones |

For the same set of references, Table XV summarizes the analysis using criteria 2 for this system. [99] integrates data fusion with EW but lacks physical scalability testing. CNN's transferability to SAR radar but with usage in fire system is evaluated in [100]. Again, radar technology is mixed with a fire direction system in [101], exploring MIMO technology in a radar using ML for overload sce-narios. [102] applies two-stage noise reduction, improving accuracy in the target direction for the weapon. [103] uses adaptive clustering for dense target extraction. [104] enhances waveform design for tactical use of radar in new fire and weapon control systems. [105] achieves real-time noise filtering with RadCNN. [106] emphasizes full SA with single data streams.

Finally, the analysis using criterion 3 is collected in Table XVI across various criteria such as war strategies, command decision support, operations, and logistics. [107] highlights material ballistic response for strategy adaptation and weapon selection, aiding in land, air, and naval

TABLE XV: Criterion 2 for Fire and Weapon Direction Systems

| | Fire and Weapon Direction System | | | | | | | | | | | |
|---|---|---|---|---|---|---|---|---|---|---|---|---|
| Ref. | Data Fusion | Tactical Scenario Inference | Command Assistance Element | Offer Roadmaps against Threats | Ergonomic Human-Machine Interface | Universal Language with Other Systems | Human Rules of Warfare | Availability | Resource Optimization | Scalability of Structures | Integrity | Reaction Speed |
| [107] | Not specified, not necessary | Not specified, not necessary. | Yes, when generating representative and additional ballistic samples for untrained classes. | Indirectly yes, as it can predict the ballistic impact of a projectile. | N/A | Recommended if linking with a weapon system. | N/A | Immediate. | Yes, when determining ballistic accuracy. | N/A | N/A | Not specified, not necessary. |
| [108] | Yes, could be combined with meltpool type images to verify in-situ hole size | Yes, 7 atmospheric parameters interfere: wind speed, pressure, temperature, humidity, dew point, solar flux, and temperature difference between air and sea. | Yes, to determine the level of damage to be caused. | Indirectly yes, by selecting attack/defense power. | Not specified as it's a proposal, but it should be ergonomic. | Recommended for linking with the weapon system. | N/A | Verified that ambient temperature is the most influential factor. | Yes, the goal is to use only the necessary power. | The algorithm could be used for electromagnetic weapons. | The distribution of ANN samples was 70% training, 15% evaluation, and 15% validation with 94.86% accuracy. | The ANN algorithm has a very high computational time cost (3 hours). RF algorithm has a relatively short training period. |
| [109] | Not specified, not necessary | Not specified, not necessary | Yes, the level the damage for each weapon | Indirectly impact to others targets | Not specified as it's a summary, but it should be ergonomic. | Not specific, but it is recommended for linking with the C2 system. | N/A | Immediate results | Not considered | Not allowed | Not specifie | Primary proposals, slow reaction |
| [110] | Without fusion, only one stream | Not considered | Not necessary | Not specified | N/A | Not specific, but it is recommended for linking with the C2 system. | N/A | Not specified | Locust Optimisation Algorithm | NN is scalable with higher layers | Not specified | Optimization reduces the search area |
| [111] | Not used | Not specified | Artillery firing data | N/A | Not considered | Not specific, but it is recommended for linking with the C2 system. | N/A | Not specified | Genetic Algorithm | Not allow | Not specified | PSO speed up the convergence of the algorithm |
| [112] | From 5 types of weapon | Not specified | Not specified | N/A | N/A | Not specific, but it is recommended for linking with the C2 system. | N/A | Not specified | Not considered | Including higher hidden layers | Accuracy level of 98%. A precision of 93.13%, recall of 94.17%, and mean values of 93.60%. | SVMs had better processing speeds. Hybrid model shows accuracy was balanced with processing speed. |

TABLE XVI: Criterion 3 for Fire and Weapon Direction Systems

| | Fire and Weapon Direction System | | | | | | | | | | | |
|---|---|---|---|---|---|---|---|---|---|---|---|---|
| Ref. | Army | War Strategies in Armed Conflicts | Command Decision Support | Cybersecurity | Military Intelligence | Unit Maintenance | New Constructions | Air Operations | Ground Operations | Naval Operations | Logistics | Unit Training |
| [107] | Land | Yes, knowing the ballistic response of materials used by both the enemy and ourselves, the attack and defense strategies change. | Yes, it helps the command select the weapon or type of ammunition (explosive, armor-piercing, proximity, etc.). | N/A. | Yes, provides information on the opponent's constructions. | N/A | Yes, it can be used as a material testing tool. | Yes | Yes | Yes | Yes, a battle scenario analysis could be performed to select the appropriate weapon and ammunition. | Yes, knowing the ballistic response of a material could train effective combat strategies. |
| [108] | Land | Yes, by consuming only the necessary energy, the unit could remain deployed for a longer time. | Yes, it helps determine the effectiveness of the weapon since it depends on factors like temperature. | N/A. | N/A. | Yes, laser power would be more regulated. | N/A | Yes | Yes | N/A | Yes, as it would affect the unit's energy consumption. | The use of this technology should be trained as a deterrent without causing direct damage. |
| [109] | Land | Any strategies is considered due to the prediction | Yes, it helps the command select the weapon or type of ammunition | N/A | N/A | Not specified, but depending of weapon size | N/A | No | Yes | No | Yes, select the best armament for battle | The prediction should be trained as a deterrent without causing direct damage. |
| [110] | Land | Any strategies is considered due to the prediction | Yes, it helps the command select the weapon or type of ammunition | N/A | N/A | Not specified, but depending of weapon size | N/A | No | Yes | No | Yes, select the best armament for battle | The prediction should be trained as a deterrent without causing direct damage. |
| [111] | Land, Naval | Not specified, any possible strategy | Yes, it helps the command select the weapon or type of ammunition | N/A | N/A | Not specified, but depending of weapon size | N/A | No | Yes | No | Yes, select the best armament of artillery | The GA algorithm should be trained as a deterrent without causing direct damage. |
| [112] | Land | Yes, the combination of different weapon within 5 categories | Yes, it helps the command select the category for the weapon. | N/A | N/A | Not specified, but depending of weapon size | N/A | No | Yes | No | Yes, select the best armament for battle | The category selection should be trained as a deterrent without causing direct damage. |

TABLE XVII: Criterion 1 for Unmanned Systems

| Ref. | Application | Objective | Innovation | AI Type | Training data information |
|---|---|---|---|---|---|
| [107] | Overcoming the mobility, communication, resource management and security challenges. | ML focused on meeting network requirements, taking into account the roles, collaboration, cooperation, and changing contexts. | Study A2A, A2G, and G2A communications to ensure QoS and QoE. | Algorithms like ANN, CNN, DNN, SVM, DQN, RandF, KNN are analyzed depending on the type of A2A, A2G, and G2A communication. | Not determined as it is an analysis and collection of different research. |
| [108] | Classification of drones using radio frequency (RF) signals. | Use of RF signals for drone detection with specific frequency ranges. | Hybrid Model with Feature Fusion Network (HMFFNet). | CNN-based feature extraction followed by feature fusion and SVM-based classification. | Features captured with a Deep Learning VGG19 network and sorting done with SVM. |
| [114] | Architectural design for automatic AV behavior generation. | Widespread and scalable decision-making framework. | Tactical and Strategic Behaviors in Automated Driving. | Behavior-Based Hierarchical Arbitration Scheme. | Database containing a merged and abstract representation of available sensor data. |

TABLE XVIII: Criterion 2 for Unmanned Systems

| Ref. | Data Fusion | Tactical Scenario Inference | Command Assistance Element | Offer Roadmaps against Threats | Ergonomic Human-Machine Interface | Universal Language with Other Systems | Human Rules of Warfare | Availability | Resource Optimization | Scalability of Structures | Integrity | Reaction Speed |
|---|---|---|---|---|---|---|---|---|---|---|---|---|
| [107] | Alternatives to ad hoc flying networks, caching or UAV processing. | Needed awareness of context changes and adaptability to current service requirements. | Yes, overcoming the 4 challenges already described. | A2A communications (participation threshold based on energy, capacity, mobility) and A2G communications (interference management and spectrum mapping). | N/A | Yes, the UAV network and base stations should be understood. | N/A | ML is a suitable solution for a dense and dynamic environment. | Yes, ML is the right tool for predicting context changes and optimization. | Yes, it should be adapted to the number of UAVs, mobility, communication, resource management and security. | Dependent on performance, communications delays and resource management efficiency. | It will depend on the functions and missions entrusted. |
| [108] | Yes, the characteristics of the 3 stages are merged for better discriminatory property. | N/A | Yes, it is another form of classification of different sound, image or radar. | N/A | Not specified but should exist. | If you want to automate the defense process. | N/A | High. | Yes, audio, image or radar sensors are not required. | N/A | Could be altered with electronic warfare and assume fake drone. | Fast |
| [114] | Merges information from all available sensors. | Contains fused, tracked and filtered representation of the world. | Yes, because AV would be autonomous. | A cost-based arbitration scheme is useful when multiple behavioral options are applied. | N/A | The same language between all sensors and the autonomous system. | It contains parking and emergency behaviors and prevents indefinite states. | Robust and efficient modular design. | Human resources would not be necessary. | Structure designed for cars in cities and roads, but could be extended to military vehicles in areas of operation. | Supports different planning approaches. | Immediate to avoid an accident. |

operations. [108] focuses on energy-efficient deployments and regulated laser power for ground operations. [109] and [110] emphasize predictive strategies for weapon selection, mainly in ground operations. [111] and [112] explore genetic algorithms and multi-category weapon combinations for artillery optimization.

### F. Unmanned Systems

Integrating ML into military drones will create a valuable weapon in armed conflicts. The ideal scenario in a land battle would be a swarm of economic, autonomous, stealthy mini-drones with enemy recognition and lethal capacity. There is still a long way to go before realizing that idea of science fiction. Still, there are already many ML studies to improve the reliability of UAVs concerning their performance and communication delays, the efficiency of resource management, and their performance based on their roles or missions, as seen in [115].

Therefore, ML has revolutionized the capabilities of US in military contexts, offering significant advantages in terms of efficiency, decision-making, and operational effectiveness. Some benefits of incorporating ML include:

- Enhancement of Autonomous Decision Making: ML algorithms enable US to process large amounts of data in real-time, allowing rapid and accurate responses to dynamic battlefield conditions. This capability reduces the reliance on human operators and improves the speed and precision of military operations.

- Predictive Maintenance: ML models can analyze data from various sensors to predict equipment failures before they occur, thus reducing downtime and maintenance costs. This predictive capability ensures that US remain operational for more extended periods, increasing their availability and reliability in critical missions.

- Operation: By integrating data from multiple sources, such as satellite imagery, radar, and on-ground sensors, ML algorithms can provide a comprehensive and coherent picture of the operational environment. This improved situational awareness is crucial for mission planning and execution, enabling more informed and effective decision-making.

- Adaptation: ML contributes to developing adaptive and resilient systems. US equipped with ML can learn from past experiences and adapt their behavior to new and unforeseen challenges. This adaptability is essential in complex and unpredictable military environments, where static programming may fall short.

TABLE XIX: Criterion 3 for Unmanned Systems

| Ref. | Army | War Strategies in Armed Conflicts | Command Decision Support | Cybersecurity | Military Intelligence | New Constructions | Air Operations | Ground Operations | Naval Operations | Logistics | Unit Training |
|---|---|---|---|---|---|---|---|---|---|---|---|
| [107] | Earth or Air and Space. | Yes, it could be a network of connected UAVs locating real-time targets. | Yes, it will depend on the use of UAVs. | It is necessary to have an autonomous defense system that guarantees the integrity, confidentiality, and availability of the data. | Yes, if it applies to recognition work. | Yes, the UAVs network study may be an inducement for new construction. | Yes | Yes | Yes | The transport of information data affects this area. | Yes, it is necessary to train the use of these networks and their defense. |
| [108] | Earth. | Yes, the radio spectrum could be analyzed to find out how many drones are in the battle. | Yes, it is a way for classifying. | Study Potential Interference. | Yes, if you can analyze the spectrum and determine no drones. | Yes, encourage to work with other forms of communication. | Yes | Yes | Yes | N/A | Yes, it could be trained in different ways to deceive the RF classification. |
| [114] | Earth. | Yes, having autonomous vehicles provides new strategies. | N/A | N/A | Yes, they could be used in recognition work. | Yes, the first 100% autonomous vehicle developments are emerging. | Yes | Yes | Yes | Yes | Yes, training is required along with autonomous vehicles. |

In summary, integrating ML into unmanned military systems offers substantial benefits, including enhanced autonomous decision-making, predictive maintenance, improved situational awareness, and adaptive capabilities. These advancements increase the effectiveness and efficiency of military operations and contribute to the safety and success of missions.

Thus, different publications aim to improve US by leveraging ML. Table XVII presents an analysis based on Criterion 1, highlighting the most relevant publications alongside the concepts discussed in the previous section and identifying the military areas they may impact focused on US. Table XVIII provides an analysis based on Criterion 2, while Table XIX focuses on Criterion 3. Note that some of the publications presented show an N/A in some criteria due to their lack of relevance.

These works highlight the advancements in US by integrating ML and improving decision-making, communication, and operational efficiency. [107] focuses on overcoming challenges in UAV mobility, communication, resource management, and security using ML techniques like ANN, CNN, DNN, SVM, and DQN to ensure QoS and Quality of Experience (QoE) in Aerial to Aerial (A2A), Aerial to Ground (A2G), and Ground to Aerial (G2A) communications. The classification of drones using radio-frequency signals with a Hybrid Model featuring a Feature Fusion Network (HMFFNet), employing CNN for feature extraction and SVM for classification, and capturing features with a DL architecture Visual Geometry Group (VGG), within this group VGG19 in a network is discussed in [108]. [114] describes an architectural design for automatic AV behavior generation, using a modular behavioral block framework for scalable decision-making, integrating tactical and strategic behaviors with a Behavior-Based Hierarchical Arbitration Scheme.

## VI. Projects and Defense Industry integrating AI

This section presents a selection of significant projects, industries, and countries related to defense where AI and ML are applied. Table ?? summarize their name, what type of initiative they are, their prominent supporters, and the core related technological systems introduced in Section III involved in them.

### A. AIDA

Thales Group has been heavily involved in incorporating AI into defense projects, focusing on enhancing military systems' security, efficiency, and autonomy. Some of their notable AI-driven defense initiatives include the Artificial Intelligence Deployable Agent (AIDA) project [116]. The European Defense Fund funded it and aims to develop an autonomous AI agent capable of enhancing cybersecurity in defense systems. Specifically, AIDA is designed to protect aircraft systems from cyberattacks, providing real-time automated threat detection and response. Thales leads the project, collaborating with multiple European partners, and the solution is tested in scenarios involving advanced cyber-electromagnetic threats and adversarial AI attacks. The project highlights Thales's strengths in onboard systems and cybersecurity, emphasizing autonomous responses to cyber threats in high-intensity environments. Thales Group is also applying AI to develop advanced radar systems for air defense [116], [117]. Thales's radar systems are designed to detect and track various aerial threats, from aircraft to missiles, in complex environments. The AI algorithms are embedded to ensure that the radar systems can autonomously adapt to different scenarios, making them more resilient to EW and capable of working in concert with other defense systems.

### B. ASTRAEA

The ASTRAEA project (Autonomous Systems Technology Related Airborne Evaluation & Assessment) [118] by BAE Systems in the United Kingdom aims to develop advanced AI and ML technologies to improve the autonomy and effectiveness of military systems, making them more capable of operating in complex and dynamic environments without constant human intervention. The project seeks to optimize the resilience of the systems so that they can continue functioning even when faced with unforeseen situations or threats.

The project focuses on the following areas:

- Integrating AI capabilities for decision-making in combat, surveillance, and logistics missions, optimizing the systems' autonomy and ability to adapt to rapid environmental changes.
- Developing technologies for autonomous air and ground vehicles that operate without direct human intervention. These systems are essential for reconnaissance, exploration, and logistical support missions in conflict zones.
- Ensuring the resilience of systems against cyberattacks and providing the ability to self-diagnose or recover from failures is a key component of the project, ensuring that systems remain uncompromised during critical missions.

As a defense and security engineering leader, BAE Systems has collaborated with various government agencies and technology companies to advance the ASTRAEA project. This includes partnerships with academic institutions and research laboratories that contribute their expertise in AI, robotics, and data analysis.

The project is part of a broader strategy by BAE Systems to innovate in the field of autonomous technology, not only developing autonomous defense systems but also seeking to improve the capabilities of armed forces by integrating new technologies into their operational structure.

### C. ATLAS

Advanced Targeting and Lethality Automated System (ATLAS) project [119] aims to provide AI and ML to U.S. combat tanks, making it possible to identify and attack three times faster than conventional procedures. For this purpose, the work has been focused on the following technology areas:

- Data collection on potential types of military targets and performing a prior training of the ML algorithm used.
- Imaging processing applying ML techniques for classification, recognition, identification, and tracking of objectives
- Shot control. In this area, advanced guiding algorithms, the automation of the shooting process, and weapon recommendations are very important to be used according to the identified objective.
- The technical support integrated into the combat vehicle due to it is necessary high voltage power system (600 Vdc) and the integration of sensors and electronics.
- Sensors. To carry out all needed automatization and provide available real data for the ML algorithm, tanks are equipped with sensors in the visible spectrum, infrared spectrum (NIR, SWIR, MWIR, and LWIR), 360º rotation of the sensors and rangefinder lasers (LADAR and LIDAR).

The ATLAS initiative harnesses the power of ML for image recognition, enabling surveillance systems to detect potential terrorist attacks and anticipate events, as outlined in [120].

### D. COBRA

The COBRA project will allow us to carry out adaptive and customizable cyber maneuvers of hyperrealistic simulation of Persistent Advanced Threats (APT) and cyber-defense training using gamification [121]. This project is a Spanish initiative based on the COINCIDENTE program of the General Directorate of Armament and Material (DGAM) of Spain [122]. It started on December 1, 2020, with a total duration of 24 months, with the collaboration of the University of Murcia, the Polytechnic University of Madrid, and the Indra company.

The main objectives of this project are presented next:

- To simulate topology networks and real traffic.
- To develop random and parameterizable scenarios.
- To develop adaptive cyber maneuvers using gamification.
- To validate the entire proposal in the Cyber Range of the Joint Cyberspace of the Ministry of Defence of Spain.

This project incorporates AI techniques with adaptive learning. The scenarios can be adapted specifically to each student and can perform adaptive cybermaneuvres with gamification. In addition, different information will be gathered through telemetry and biometric systems.

In addition, ML enables the system to recognize the visual shape of an enemy tank, detect its thermal signature, and establish alarm parameters. When satellite images capture a figure resembling these characteristics, the system promptly alerts the operator. This approach reduces reliance on the sensitivity of surveillance personnel while significantly expanding the monitored area, enhancing both efficiency and coverage.

### E. DARPA

Defense Advanced Research Projects Agency (DARPA) [123] is an agency of the United States Department of Defense responsible for research and development of new technologies and innovative systems to develop disruptive technologies that can transform the way armed forces operate, giving the United States a technological advantage on the battlefield. DARPA includes several applied projects integrating AI into defense.

- OFFensive Swarm-Enabled Tactics (OFFSET) [124]: Aims to develop swarms of small autonomous drones capable of operating together to perform reconnaissance, attacks, and rescue missions.
- Lifelong Learning Machines (L2M) [125]: Seeks to use ML to train cybersecurity systems capable of detecting threats and continuously adapting to new attack tactics. Creating autonomous systems that can defend computer networks and protect critical infrastructures against cyber threats.
- AI for Military Operations (AIMO) [126]: A program aimed at developing AI technologies that help improve the precision of military operations and optimize resources. The project also addresses how to efficiently integrate AI into joint military operations, where different branches of the armed forces work in coordination.

### F. General Dynamics

General Dynamics is another major defense company in the United States. General Dynamics is using AI in the development of armored and combat vehicle systems, as well as in enhancing real-time intelligence capabilities for the armed forces. Through its unit General Dynamics Land Systems, the company has been developing autonomous armored vehicles for the U.S. Army under the Robotic Combat Vehicle (RCV) program. These vehicles are designed to operate without direct human intervention and perform tasks such as reconnaissance, target attacks, and logistical support on the battlefield. The incorporation of ML algorithms aims to improve autonomous driving, navigation, and decision-making in armored vehicles [127], [128].

On the other hand, it has started expanding its developments into cybersecurity systems, using AI to detect and neutralize cyber threats.

### G. GIDE

Global Information Dominance Experiment (GIDE) project is aimed to predict possible threats using AI and ML to analyze the information provided by satellites, radars, drones, underwater capabilities, networks, and others [129]. This technology would allow the U.S. military to view movements several days before the enemy, providing an advantageous tactical environment over any attack.

However, the technology used in this project is not novel. Innovative is using AI and ML to change how information and data are used. ML and AI allow a set of different parameter alert configurations and perform tests with another kind of Geospatial Intelligence (GEOINT) sensors to closely observe what is happening at a specific location [130].

### H. Iron Dome

Israel has led the implementation of advanced technologies, including AI, in its defense system. This includes using AI for threat prediction, intelligence data analysis, and the enhancement of missile systems. A notable example is the Iron Dome project [131], an air defense system developed by Israel to intercept and destroy short-range missiles, rockets, and artillery shells that threaten civilian areas. Developed by Rafael Advanced Defense Systems and Israel Aerospace Industries, the system has proven highly effective in protecting Israeli populations from aerial attacks from Gaza and other regions.

Its operation involves the following key components:

- Detection radar: The system uses advanced radars to detect real-time threats such as incoming rockets and missiles. These radars provide high-precision data about the trajectory of the projectiles.
- Battle control center: Once the threat is detected, the system performs an automatic analysis to determine if the projectile is a real threat to the protected areas. If the projectile is deemed capable of causing damage, the system autonomously intercepts it.
- Interceptors: The Iron Dome interceptors are launched to destroy the incoming projectile in the air before it can reach its target. The system has a high success rate, intercepting more than 90% of threats aimed at civilian areas.

## I. Lockheed Martin

Lockheed Martin is one of the largest and most prominent companies in the defense and aerospace sector, based in Bethesda, Maryland, USA. It manufactures some of the world's most advanced stealth combat aircraft in the defense sector, such as the F-22 Raptor and the F-35 Lightning II [132]. It also develops various missile systems, including the Terminal High Altitude Area Defense and Patriot missile defense systems to intercept ballistic and cruise missiles.

Involved in the current development of advanced technologies and innovative projects, Lockheed Martin has been utilizing AI and ML in several aspects of defense, including updates to the F-35 and intelligent missile systems. Additionally, they implement AI in predictive maintenance and data analysis to enhance cybersecurity and space defense through:

- Autonomy in aircraft and UAVs: The use of AI for real-time decision-making during combat or reconnaissance missions
- AI for failure prediction: They use predictive models to anticipate failures in system components, optimizing maintenance.

## J. Maven

Project Maven [133], initiated by the U.S. Department of Defense, focuses on applying AI and ML to analyze drone footage to identify and track objects of interest, such as potential targets. This project began in 2017 and uses computer vision algorithms to sift through massive amounts of video data gathered by drones, significantly improving the speed and accuracy of target recognition compared to manual methods. The primary goal of Project Maven is to assist the military in making faster, more informed decisions by automating the analysis of vast amounts of surveillance footage, enabling a more efficient use of resources in intelligence gathering and combat operations.

The AI-powered system processes video feeds to detect patterns and identify objects, which can be crucial in military operations such as targeting, reconnaissance, and surveillance. This system can identify vehicles, people, or other objects and provide real-time data for further action. Despite concerns over the ethics of AI in warfare, particularly regarding the automation of target identification, Project Maven has sparked significant interest in integrating AI into military applications.

## K. NORINCO

China has been investing significantly in AI to enhance its defense capabilities, particularly cybersecurity and military automation. The country has implemented AI in various systems for mass surveillance, troop control, and military vehicle automation. One prominent focus is cyber defense, where AI analyzes vast amounts of intelligence data, detects cyber threats, and improves national security. Additionally, AI is being used to develop autonomous military systems, including drones and unmanned vehicles, designed to carry out combat and reconnaissance missions with minimal human intervention. These technologies are central to China's vision of advancing its military capabilities through AI, emphasizing automation and rapid decision-making.

In this context, China North Industries Group Corporation (NORINCO) [134] is one of the key players actively developing AI-powered drones and autonomous military vehicles for defense applications. A notable example is their anti-drone technology [135], part of their larger EW systems for armored vehicles, particularly their VT4A main battle tanks. These systems utilize AI and radar technology to detect, track, and neutralize threats from small, slow-moving drones. These AI-driven systems provide layered defense strategies for ground units, enhancing the capabilities of military platforms to defend against modern drone threats. NORINCO also showcased these technologies at the Airshow China 2024, emphasizing the shift toward more digitally empowered and adaptable defense solutions in response to evolving combat scenarios

## L. Northrop Grumman

Northrop Grumman is a leading defense and aerospace technology company recognized for its innovation in applying AI and ML to modern military systems [136], [137]. The company integrates AI across various defense sectors, including space, cybersecurity, and autonomous vehicles. One of the most significant uses of AI by Northrop Grumman is in satellite defense systems, where AI is employed to monitor and protect satellites from potential threats, enhancing space-based security. Additionally, the company has been developing autonomous drones that utilize AI for real-time decision-making during combat and reconnaissance missions. These AI-driven drones aim to increase operational efficiency and reduce human intervention in high-risk environments.

One of Northrop Grumman's ongoing initiatives involves leveraging AI in advanced radar systems and missile defense technologies, including predictive algorithms that improve defense systems' accuracy and response time. This effort aims to increase the precision of military operations, provide faster and more reliable defense mechanisms, and optimize resource allocation during missions.

Northrop Grumman's contributions to AI in defense align with broader trends in the defense industry, where AI is becoming an integral part of decision-making, autonomous systems, and cybersecurity.

## M. Russia

Russia has been increasingly focused on advancing AI technologies to enhance its defense capabilities. The country invests heavily in autonomous robotics, EW, and intelligent missile systems. Integrating AI into military systems aims to improve operational efficiency, precision, and adaptability in dynamic combat environments.

One of Russia's prominent projects involves the development of autonomous combat robots. These include ground-based robots and combat drones that operate independently, leveraging AI algorithms to perform reconnaissance, target identification, and attack operations. AI in these robots allows them to operate in highly complex and unpredictable environments, reducing human risk and increasing combat effectiveness. The Uran-9 combat robot, for example, is a key system developed by Russia that is designed to operate autonomously in combat zones, showcasing the country's ambition to integrate AI into its military assets.

Another critical development area is intelligent missile systems incorporating AI to enhance targeting precision and adaptability to changing conditions. Russia's use of AI in missile technology improves accuracy, allowing these weapons to adjust in real-time to counter defensive measures or alter course in unpredictable battlefield scenarios. This is expected to significantly increase the effectiveness of missile strikes, even in complex and rapidly changing operational environments.

In addition to these projects, Russia has been focusing on EW systems that use AI to detect and counter adversary signals, disrupt communications, and neutralize enemy systems in the electromagnetic spectrum.

Russian Defense Companies Involved:

- Kalashnikov Group: known for developing autonomous combat systems, including AI-powered drones and robots [138].
- Almaz-Antey, a defense manufacturer that works on advanced missile systems and air defense solutions that integrate AI for more efficient target acquisition and defense [139].

These developments position Russia as a key player in the evolving AI-driven defense sector, emphasizing autonomy, intelligence, and adaptability in warfare.

### N. SEDA

The SatEllite Data Ai (SEDA) project [140], [141], a Spanish initiative based on the COINCIDENTE program of the General Directorate of Armament and Material (DGAM) of Spain [122], emerged at the end of 2018, an intelligent geospatial that analyzes and exploits satellite information to detect changes in the temporal status of satellite imagery.

This project combines the potential of DL with data processing and data fusion advances to analyze satellite information automatically. The resulting tool allows the discovery of information that is not fully revealed, such as the movement of troops or war equipment.

### O. SOPRENE

SOPRENE (Predictive Sustainment of Neural Networks, or SOstenimiento Predictivo de REdes NEuronales in Spanish) is a Spanish initiative based on the COINCIDENTE program of the General Directorate of Armament and Material (DGAM) of Spain [122] that promotes the use of neural networks in the preventive maintenance of Spanish Navy ships [142].

The most modern ships in the Spanish Navy have installed an integrated platform control system, which mainly controls their propellant plant, power plant, auxiliary machines, and firefighting system. This platform or system is also associated with a Condition Based Maintenance System. Hundreds of ship sensors are connected, generating thousands of signals and hundreds of megabytes of daily information. This data is sent to the Navy Data Monitoring and Analysis Center to create a Big Data signal architecture of propulsion motors, electric generators, fire pumps, and other equipment to analyze and predict faults or abnormal performances.

### VII. Synthesis of Results from analysis

Once the literature has been reviewed to identify works that propose techniques from AI, the ML family, or the DL group, we analyze the data to draw statistical conclusions and identify gaps in the methods used in the defense sector. First, we have compiled in Table XX the techniques used so far for each type of learning. The histogram in the Fig. 14 shows the distribution of techniques used in the defense sector. Overall, it can be observed that three predominant techniques stand out in defense technologies: CNN, RF, and SVM. These are also well-established and widely used in the civilian sector, further reinforcing the argument presented in this study regarding the potential for cross-domain adoption and reuse of AI techniques in both military and civilian applications.

TABLE XX: AI classifications per learning type

| Category | Algorithms/Techniques |
|---|---|
| Supervised ML | SVM, Bayes Point Match, Boosted Decision Trees, Decision Forests, Decision Jungles, Native Bayes Classifier, Gradient Boosting Machine, Distributed Random Forest, Logistic Regression, XGBoost, CNN-SVM |
| Non-supervised ML | K-Means, Non-ML Optimization, CL techniques |
| Reinforcement Learning | LSTM-DQN, Hybrid AI model, DRL, DDPG |
| Deep Learning | DNN, ANN, RNN, GNN, RadCNN, VGG16, YOLOv3, YOLOv4, YOLOv7, YOLOv8, GAN, Mask R-CNN, CNN, ResNet50, Xception, Siamese-CNN, cGAN, GOA-RNN, GA-LFPSO |

Meanwhile, Fig. 15 represents the data in percentage form, considering the 110 AI studies applied to military systems. Analyzing by learning type, Fig. 15 illustrates the distribution of AI learning paradigms applied in military systems, highlighting the predominance of DL with 37% of the techniques used. Its effectiveness in image recognition, radar analysis, and autonomous decision-making drives this dominance. SL follows with 24%, playing a crucial role in classification tasks, predictive maintenance, and electronic warfare signal processing. RL and Hybrid AI Models each account for 15%, reflecting their growing importance in adaptive decision-making and multi-model integration for defense applications. UL comprises 8%, primarily used in clustering and data optimization where labeled data is scarce. Finally, Federated Learning (FL) represents 1%, indicating its emerging but still limited adoption in military AI, likely due to operational constraints and security concerns.

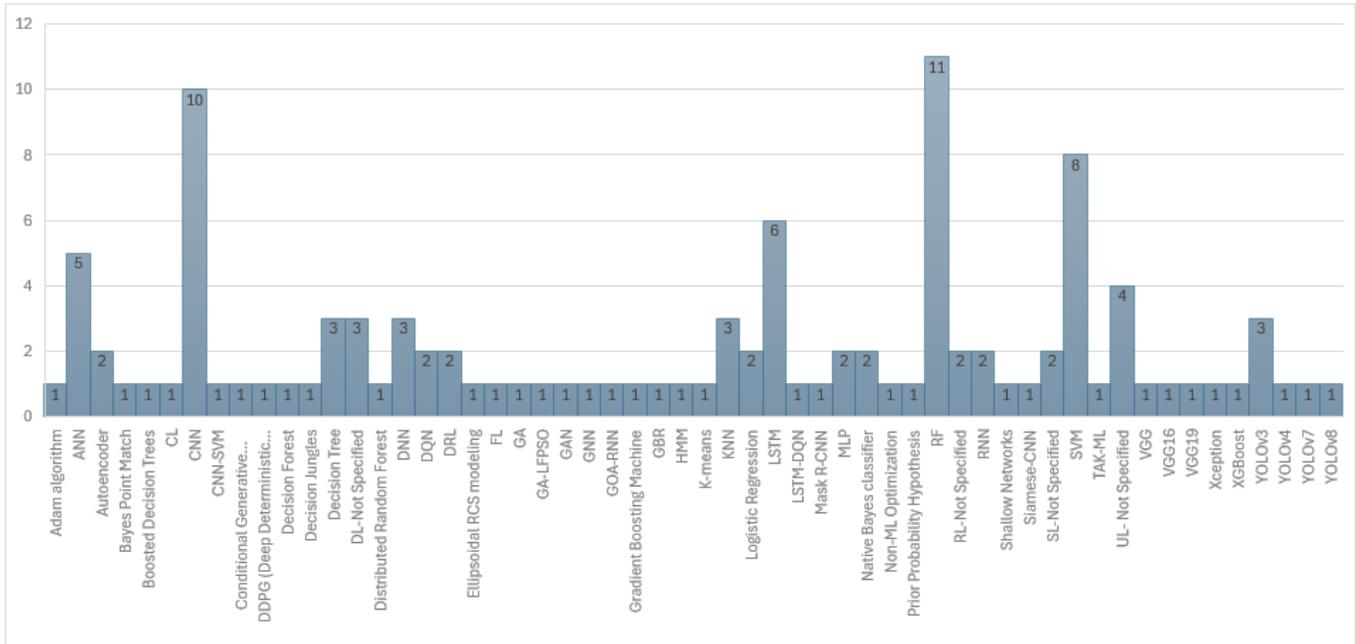

Fig. 14: Mapping between Number of references and type of AI, ML, DL used.

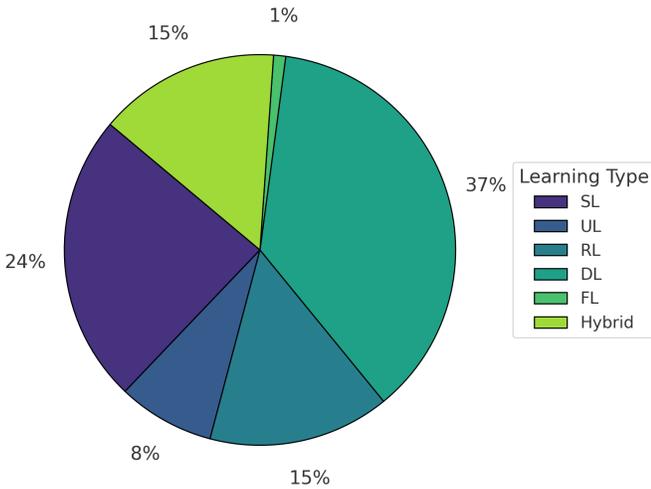

Fig. 15: Statistical per learning type.

A more detailed analysis reveals key trends in adopting and distributing different AI technologies across defense applications. These trends are shaping the future of defense capabilities, with each system leveraging specific AI tailored to its operational requirements.

The stacked bar chart in Fig. 16 illustrates the distribution of AI usage in military systems. DL, including CNN, DNN, YOLO, and GAN models, is the most utilized technology, particularly in Image Surveillance and Radar Systems, where object recognition and threat detection are crucial. Reinforcement Learning (DRL, DQN) is predominantly applied in Tactical Information Systems and Unmanned Systems, enhancing autonomous decision-making and adaptability in complex environments. Supervised Learning (SVM, RF) is widely employed in EW and Predictive Maintenance, ensuring efficient classification and fault detection. Evolutionary Algorithms (GA, PSO) are mainly integrated into Fire and Weapon Control Systems to optimize targeting accuracy and trajectory prediction. At the same time, Hybrid AI Models exhibit a more balanced but lower frequency of application across different military domains.

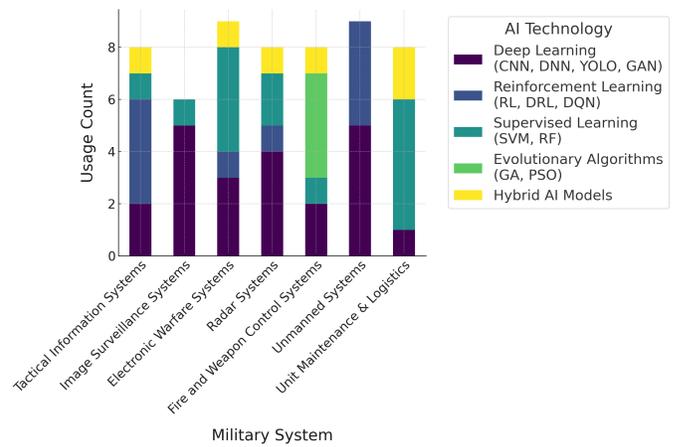

Fig. 16: Distribution Of AI Technologies Across Military Systems

A trend analysis in Fig. 17 highlights the increasing prevalence of DL as the dominant AI technology in military applications, demonstrating its effectiveness in image analysis, surveillance, and sensor-based intelligence. RL is emerging as a crucial tool for real-time decision-making and adaptive military strategies, particularly in automated combat and UAV operations. SL plays a significant role in cybersecurity, EW, and predictive logistics. Evolutionary Algorithms and Hybrid AI Models, while less widespread, remain essential for specialized applications such as ballistic computations and real-time optimization.

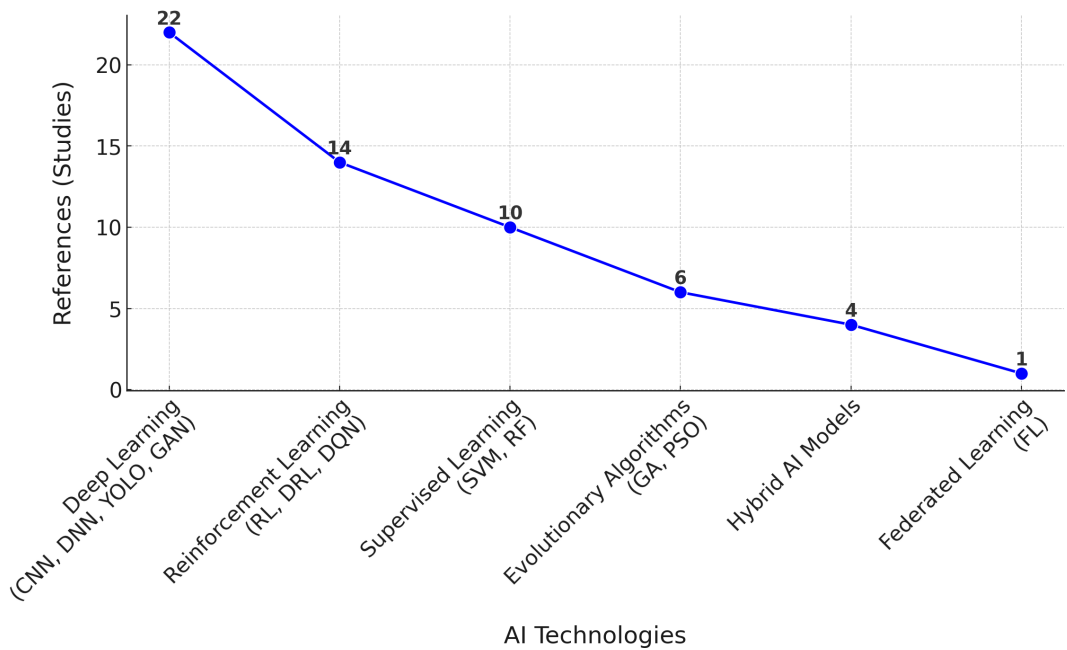

Fig. 17: Trend of AI usage in military systems

Key observations indicate that Image Surveillance and Radar Systems are leading in AI integration, mainly through advanced DL techniques. Maintenance and Logistics Systems rely heavily on SL to optimize operational efficiency and failure prediction. UAVs and autonomous vehicles are evolving with RL, enhancing their ability to adapt to dynamic environments. The application of AI in Electronic Warfare and Fire Control Systems still requires further advancements in precision and adversarial resilience. These insights suggest that future research should focus on improving interoperability among AI-driven military technologies and leveraging Reinforcement Learning and Deep Learning to strengthen autonomous defense and electronic warfare capabilities.

These results confirm the widespread adoption of DL and SL in military AI, closely mirroring their prevalence in civilian applications. Meanwhile, RL and Hybrid AI are gaining traction, particularly in autonomous systems and strategic operations. The minimal presence of FL suggests a potential area for future research, emphasizing the need for secure and decentralized AI solutions in defense.

These advancements are actively being implemented in major military projects, including Project Maven, GIDE, ATLAS, DARPA OFFSET, Iron Dome, ASTRAEA, and SOPRENE. The synergy between AI and defense technologies fosters a new era of automation, efficiency, and strategic superiority.

Future research should focus on improving interoperability, data security, adversarial robustness, and the ethical deployment of AI in military environments. Bridging the gap between civilian and military AI developments through cross-domain AI integration, federated learning, and secure collaboration will ensure AI-driven military innovations remain ethical, accountable, and strategically viable.

## VIII. Future Research Directions

Integrating AI into military communications and networking presents significant challenges and opportunities for the future research lines. While AI has demonstrated potential in areas such as decision-making, autonomous systems, and electronic warfare, its full deployment in tactical communication networks and multi-domain information sharing requires further advancements. This section explores key challenges, emerging trends, and strategic recommendations to guide future research and innovation.

### A. Key Challenges in AI-Driven Military Communications

- Interoperability Across Multi-Domain Operations: AI-driven military networks must integrate seamlessly across land, air, sea, space, and cyber domains. Current AI models often lack the adaptability required to function efficiently in diverse operational environments, necessitating the development of standardized communication protocols.

- Adversarial Threats, Cybersecurity Risks, and Ethical Considerations: AI-enabled defense networks are vulnerable to EW, cyberattacks, and adversarial AI tactics. Future research must focus on robust AI security mechanisms, encryption-enhanced network resilience, and real-time threat detection algorithms. Ethical concerns surrounding AI in warfare require standardized frameworks and regulations to ensure accountability and compliance.
- Data Scarcity and Real-World Adaptability: Military AI models require extensive, high-quality data to enhance learning capabilities. However, access to real-world datasets is restricted due to security concerns. The development of synthetic data generation and simulation-based AI training is crucial to overcoming these limitations.
- Scalability and Latency in Tactical Communications: AI-driven communication systems must operate in low-latency, high-mobility battlefield environments. Future research should focus on optimizing real-time AI inference, edge computing for tactical units, and decentralized AI architectures to reduce reliance on centralized cloud processing.

B. Emerging Trends in AI for Military Communication Networks

- FL for Secure AI Model Training: Distributed AI training allows allied nations and defense units to develop AI models collaboratively without sharing raw data, improving confidentiality while enhancing AI performance.
- AI-Enhanced Network Resilience and Self-Healing Communications: AI-based autonomous recovery mechanisms can enhance the survivability of battlefield networks by dynamically reconfiguring communication pathways in response to disruptions.
- Cognitive Radio and AI-Driven Spectrum Management: AI can optimize spectrum allocation, interference mitigation, and adaptive frequency hopping to ensure uninterrupted military communications in congested or adversarial environments.
- Neuromorphic Computing for Low-Power AI in Tactical Networks: Neuromorphic computing enables low-power, high-efficiency AI models for real-time signal processing and decision-making in battlefield networks. These architectures offer advantages for on-device intelligence in edge computing environments, allowing UAVs, UGVs, and remote sensors to operate with minimal latency and reduced energy consumption. Future research should explore neuromorphic chips for adaptive AI models in contested electromagnetic environments, ensuring robustness in military operations.

C. Strategic Recommendations for Future AI Research in Defense Communications

- Developing Standardized AI Interoperability Frameworks: Establishing unified protocols for AI-driven communication systems will enable seamless integration between different military branches and allied forces.
- Enhancing AI Resilience Against EW and Cyber Threats: Research should prioritize adversarial training, AI-driven jamming detection, and AI-based cyber deception strategies to counteract evolving threats.
- Investment in AI-Powered Tactical Edge Computing: The deployment of AI-enabled edge devices will enhance battlefield decision-making capabilities while reducing dependence on centralized computing infrastructures.
- Advancing AI-Driven UAV and UGV Communications for Tactical Networks: While AI-powered UAVs and UGVs have been successfully deployed for tactical communications, challenges remain in optimizing their network coordination, adaptability, and resilience in contested environments. Future research should focus on enhancing real-time adaptive routing, developing self-learning communication protocols, and integrating AI-based dynamic spectrum allocation to improve interoperability across multi-domain operations. Additionally, advancements in federated learning and neuromorphic computing could enable greater autonomy and efficiency in UAV-UGV communication networks, ensuring seamless integration with existing defense infrastructure.
- Strengthening Civilian-Military AI Collaboration: Encouraging partnerships between defense agencies, academic institutions, and private AI developers can accelerate innovation in military communication technologies.

By addressing these research challenges and advancing AI technologies in tactical communications and networking, military operations can achieve enhanced security, efficiency, and resilience. Future work must focus on bridging gaps between AI research and real-world deployment, ensuring that AI-driven defense networks remain adaptable, secure, and mission-ready.

IX. Conclusion

This survey analyzes AI applications in tactical communications and networking, highlighting advancements, challenges, and gaps. While AI is growing in cybersecurity and intelligence, its use in logistics, electronic warfare, radar, information systems, and battlefield decision-making is still limited.

We emphasize the need for cross-domain AI integration across land, sea, air, and space operations. Simulation-based AI testing is vital for ensuring model reliability before deployment. Ethical concerns and adversarial threats require standardized frameworks and strong countermeasures against AI vulnerabilities. This study also stresses civilian-military collaboration to enhance innovation and draw on advancements from research, industry, and academia.

This survey is a valuable resource for researchers, military professionals, and policymakers. It provides an overview of AI's evolving role in defense. It identifies research gaps, recommends strategies, and discusses emerging trends to inform future AI-driven tactical communications developments. Success hinges on collaboration, secure data sharing among allies, and investment in resilient AI systems. Tackling these challenges will empower AI to improve efficiency, decision-making, and autonomy, ensuring security and accountability.

## Acronyms

| | |
|---|---|
| 5G | 5th Generation |
| A2A | Aerial to Aerial |
| A2G | Aerial to Ground |
| AE | AutoEncoder |
| AESA | Active Electronically Scanned Array |
| AI | Artificial Intelligence |
| AIDA | Artificial Intelligence Deployable Agent |
| AMDR | Air and Missile Defense Radar |
| ANN | Artificial Neural Networks |
| ATLAS | Advanced Targeting and Lethality Automated System |
| C2 | Command and Control |
| C2IS | Command and Control Information Systems |
| C4ISR | Command Control Communications Computers Intelligence Surveillance and Reconnaissance |
| CCD | Charge-Coupled Device |
| CCTV | closed-circuit television |
| CEW | Cognitive Electronic Warfare |
| CFAR | Constant False Alarm Rate |
| cGAN | Conditional Generative Adversarial Network |
| CL | Cooperative Learning |
| CNN | Convolutional Neural Network |
| COF | Cost Objective Function |
| CSI | Channel State Information |
| DARPA | Defense Advanced Research Projects Agency |
| DIL | Disconnected Intermittent and Limited |
| DL | Deep Learning |
| DNN | Deep Neural Networks |
| DQN | Deep Q-Network |
| DRL | Deep Reinforcement Learning |
| DSG | Deep Supervised Generative |
| DT | Digital Twin |
| ECM | Electronic Countermeasures |
| EPM | Electronic Protective Measures |
| ESM | Electronic Support Measures |
| EW | Electronic Warfare |
| FL | Federated Learning |
| FSO | Free Space Optics |
| FWDS | Fire and Weapon Direction System |
| G2A | Ground to Aerial |
| GA | Genetic algorithms |
| GAN | Generative Adversarial Network |
| GaN | Gallium Nitride |
| GEOINT | Geospatial Intelligence |
| GIDE | Global Information Dominance Experiment |
| GNN | Graph Neural Network |
| GNSS | Global Navigation Satellite System |
| HMFFNet | Hybrid Model featuring a Feature Fusion Network |
| IHL | International Humanitarian Law |
| IoBT | Internet of Battlefield Things |
| IoT | Internet of Things |
| IR | Infra Red |
| ISR | Intelligence Surveillance and Reconnaissance |
| JREAP | Joint Range Extension Application Protocol |
| KNN | K-Nearest Neighbors |
| L2M | Lifelong Learning Machines |
| LFPSO | Levy Flight Particle Swarm Optimization |
| LLM | Large Language Model |
| LSTM | Long Short-Term Memory |
| LWIR | Long-Wave Infrared |
| mAP | mean Average Precision |
| MIMO | Multiple-Input Multiple-Output |
| ML | Machine Learning |
| MLOp | ML Operations |
| MLP | Multi-Layer Perceptron |
| MNN | Multi-Layer Neural Networks |
| MWIR | Mid-Wave Infrared |
| NATO | North Atlantic Treaty Organization |
| NIR | Near Infrared |
| PSO | particle swarm optimization |
| QoE | Quality of Experience |
| QoS | Quality of Service |
| RCNN | Region-based Convolutional Neural Network |
| RCS | Radar Cross Section |
| RCV | Robotic Combat Vehicle |
| RF | Random Forest |
| RIS | Reconfigurable Intelligent Surfaces |
| RL | Reinforcement Learning |
| RMA | Radar Modular Assemble |
| RNN | Recurrent Neural Network |
| ROI | Region of Interest |
| SA | situational awareness |
| SAR | Synthetic Aperture Radar |
| SDR | Software Defined Radio |
| SEDA | SatEllite Data Ai |
| SL | Supervised Learning |
| SNR | Signal-Noise Ratio |
| SSL | Semi-supervised Learning |
| SVM | Support Vector Machines |
| SWIR | Short-Wave Infrared |
| TAK-ML | Tactical Assault Kit-ML |
| U.S. | United States |
| UAS | Unmanned Aerial System |
| UAV | Unmanned Aerial Vehicle |
| UGV | Unmanned Ground Vehicle |
| UL | Unsupervised Learning |
| US | Unmanned Systems |
| USV | Unmanned Ship Vehicle |
| VGG | Visual Geometry Group |
| VMF | Variable Message Format |
| YOLO | You Only Look Once |